\documentclass{aa}
\usepackage{natbib}
\bibpunct{(}{)}{;}{a}{}{,}
\graphicspath{{./}{figures/}}

\newcommand\swx{{\it Swift-{\it XRT}}}
\newcommand\swb{{\it Swift-{\it BAT}}}
\newcommand\cha{\textit{Chandra}}
\newcommand\XMM{{XMM-{\it Newton}}}
\newcommand\NuSTAR{{\it NuSTAR}}
\newcommand\XSPEC{{\tt XSPEC}}
\newcommand\mekal{{\tt mekal}}
\newcommand\MYTorus{{\tt MYTorus}}
\newcommand\borus{{\tt borus02}}
\newcommand\bntorus{{\tt BNtorus}}

\usepackage{courier}
\usepackage[T1]{fontenc}
\usepackage{mathtools}
\usepackage{changepage} 
\usepackage{url,multirow}
\usepackage{appendix}
\usepackage{microtype}
\DeclareUnicodeCharacter{2212}{-}
\usepackage{lineno}
\usepackage{textcomp}

\usepackage[unicode=true,pdfusetitle,
 bookmarks=true,bookmarksnumbered=false,bookmarksopen=false,
 breaklinks=false,pdfborder={0 0 0},pdfborderstyle={},backref=false,colorlinks=true]
 {hyperref}
\hypersetup{
 urlcolor=blue,citecolor=blue}

\usepackage{xcolor}
\usepackage{scrextend}
\usepackage{amsmath}
\usepackage{algorithm}
\usepackage{amsfonts}
\usepackage{mathrsfs}
\usepackage{bm}
\usepackage{rotating}
\usepackage{color}
\usepackage{graphicx}
\usepackage{subfigure}
\usepackage{float}
\usepackage{gensymb}
\usepackage{soul}

\makeatletter
\renewcommand*\aa@pageof{, page \thepage{} of \pageref*{LastPage}}
\makeatother

\begin{document}

   \title{Compton-thick AGN in the NuSTAR Era X: Analysing seven local CT-AGN candidates}

    \author{D. Sengupta\inst{1,2}\thanks{\protect\href{mailto:mailto:phydhrubo@gmail.com}{phydhrubo@gmail.com}}
            \and S. Marchesi\inst{1,2,3}\thanks{\protect\href{mailto:mailto:stefano.marchesi@inaf.it}{stefano.marchesi@inaf.it}}
            \and C. Vignali\inst{1,2}
            \and N. Torres-Albà\inst{3}
            \and E. Bertola\inst{1,2}
            \and A. Pizzetti\inst{3}
            \and G. Lanzuisi\inst{2}
            \and F. Salvestrini\inst{4}
            \and X. Zhao\inst{5}
            \and M. Gaspari\inst{10}
            \and R. Gilli\inst{2}
            \and A. Comastri\inst{2} 
            \and A. Traina\inst{1,2}
            \and F. Tombesi\inst{6,7,8,9}
            \and R. Silver\inst{3}
            \and F. Pozzi\inst{1}
            \and M. Ajello\inst{3}}
    \institute{Dipartimento di Fisica e Astronomia (DIFA), Università di Bologna, via Gobetti 93/2, I-40129 Bologna, Italy
               \and INAF-Osservatorio di Astrofisica e Scienza dello Spazio (OAS), via Gobetti 93/3, I-40129 Bologna, Italy
               \and Department of Physics and Astronomy, Clemson University, Kinard Lab of Physics, Clemson, SC 29634, USA
               \and INAF – Osservatorio Astrofisico di Arcetri, Largo E. Fermi 5, 50125, Firenze, Italy
               \and Center for Astrophysics, Harvard \& Smithsonian, 60 Garden Street, Cambridge, MA 02138, USA
               \and Department of Physics, Tor Vergata University of Rome, Via della Ricerca Scientifica 1, 00133 Rome, Italy
               \and INAF – Astronomical Observatory of Rome, Via Frascati 33, 00040 Monte Porzio Catone, Italy
               \and Department of Astronomy, University of Maryland, College Park, MD 20742, USA
               \and NASA Goddard Space Flight Center, Greenbelt, MD 20771, USA
               \and Department of Astrophysical Sciences, Princeton University, Princeton, NJ 08544, USA}

\titlerunning{Analysing seven local CT-AGN candidates}
\authorrunning{Sengupta et al.}

\abstract
    {We present the broad-band X-ray spectral analysis (0.6-50 keV) of seven Compton-thick active galactic nuclei (CT-AGN; line-of-sight, l.o.s., column density $>10^{24}$ cm$^{-2}$) candidates selected from the \swb\ 100-month catalog, using archival \NuSTAR\ data.}{We aim at obtaining a complete census of the heavily obscured active galactic nuclei in the local Universe ($z<0.05$).}{This work is in continuation of the on-going research of the Clemson-INAF group to classify CT-AGN candidates at redshift $z<0.05$, using physically-motivated torus models.}{Our results confirm that three out of seven targets are \textit{bona-fide} CT-AGN. Adding our results to the previously analysed sources using \NuSTAR\ data, we increase the population of \textit{bona-fide} CT-AGN by $\sim9\%$, bringing the total number to 35 out of 414 AGN. We also performed a comparative study using \MYTorus\ and \borus\ on the spectra in our sample, finding that both physical models are strongly consistent in the parameter space of l.o.s. column density and photon index. Furthermore, the clumpiness of the torus clouds is also investigated by separately computing the l.o.s. and average torus column densities, in each of the seven sources. Adding our results to all the previous 48 CT-AGN candidates analysed by the Clemson-INAF research team having \NuSTAR\ observations: we find that $78\%$ of the sources are likely to have a clumpy distribution of the obscuring material surrounding the accreting supermassive black hole.}{}


\maketitle

\section{Introduction}\label{sec:intro}

Diffuse X-ray emission from the central regions of accreting supermassive black holes in active galactic nuclei (AGN) is responsible for most of the cosmic X-ray background (CXB) radiation from a few keV to a few hundred keV \citep{comastri2004compton, gilli2007synthesis, ueda2014toward}. The contribution of unobscured AGN to the CXB is almost completely resolved into point-like sources at E$<10$ keV \citep{worsley2005unresolved, Hickox2006}. CT-AGN (CT-AGN; line-of-sight, l.o.s., column density $>10^{24}$ cm$^{-2}$) significantly contribute ($\sim15-30\%$ \citealt{gilli2007synthesis,ananna19synthesis}) to the CXB around its peak ($\sim20-30$ keV; \citealt{ajello2008cosmic}). In the local universe ($z\leq0.1$), the fraction of CT-AGN revealed by the X-ray observations is found to be $\sim5-10\%$ (\citealt{Vasudevan2013, Ricci_15_CTAGN-candidates,Torres2021_CT-fraction}). This reveals a big gap with the predictions of AGN population synthesis models, which postulates that the fraction of local CT-AGN should be of $\sim20-50\%$ (\citealt{ueda2014toward,ananna19synthesis}) to model the CXB properly.

For low-redshift AGN, the circum-nuclear dusty torus clouds are considered as the dominant medium of obscuration, i.e. obscuration from the inter-stellar medium (ISM) of the host galaxy is expected to be less significant (e.g. \citealt{Gilli2022}). Due to significant suppression of intrinsic X-rays below 10 keV by these obscuring Compton-Thick clouds, it is difficult to detect heavily obscured AGN at $z\sim0$ in the soft X-ray (E$<10$ keV) band. Since heavily obscured AGN have a noticeable Compton hump at $\sim20-40$ keV, hard X-ray (E$>10$ keV) observatories allow for the detection and characterization of this kind of sources at $z\sim0$. For example, the \textit{Swift} Burst Alert Telescope (BAT) is used as it is less biased against CT-AGN sources, being sensitive in the 15-150 keV range. To determine the existing CT-AGN fraction, using a BAT volume-limited sample is among the most efficient ways to reduce the bias against the obscured sources. The importance of an obscuring medium surrounding the meso-scale ($\sim1-100$ pc) around AGN has been highlighted by several theoretical and numerical investigations (\citealt{Gaspari2020}, for a review). In a nutshell, in realistic turbulent environments, the host diffuse medium is expected to recurrently condense in a top-down multiphase condensation cascade of warm and cold clouds, which then rain onto the central AGN. Such chaotic cold accretion (CCA; \citealt{Gaspari2013}) is thus often responsible for a clumpy distribution at the meso-scale, as well as boosting the feeding rates at the micro scale ($<1$ pc). This multi-scale rain has been constrained and detected in a wide range of galaxies and AGN (e.g., \citealt{Rose2019,Gaspari2019,Maccagni2021,Marchesi2022,McKinley2022,Temi2022}). 

The 100 month Swift-BAT catalog\footnote{\url{http://bat.ifc.inaf.it/100m_bat_catalog/100m_bat_catalog_v0.0.htm}} (the updated 150 month BAT catalog is in preparation, K. Imam et al.) consists of 414 AGN at $z<0.05$. From this AGN population, our Clemson-INAF research group\footnote{\url{https://science.clemson.edu/ctagn/}} have selected a sample of 55 CT-AGN candidates which have archival and Guest Observer observations with the Nuclear Spectroscopic Telescope Array (\NuSTAR\/; \citealt{Harrison2013Nustar}). The target sources are observed by \NuSTAR\ within the energy range of 3-79 keV, with high sensitivity, since \NuSTAR\ is the first instrument to focus X-ray photons at E$>$10 keV. For the soft X-ray coverage at E$<$10 keV, we have used the available X-ray spectra from \XMM\/, \cha\/ or \swx\/. We have carried out a systematic and comprehensive spectral analysis in $0.6-50$ keV band, on each of the 55 sources by using the uniform torus models- \MYTorus\ and \borus\/ (See \cite{marchesi2018compton,Zhao2021_NustarAGN,Torres2021_CT-fraction,Traina2021}). In this work, we are showing the results of the last seven sources from this sample. Here, we have independently computed line-of-sight (l.o.s.) column density (N$_{H,l.o.s.}$) and average torus column density (N$_{H,avr}$ or N$_{H,tor}$), to study the clumpiness of the torus clouds even within the uniform torus framework. The N$_{H,l.o.s.}$ is derived from the absorbed powerlaw coming directly from the `corona'. The N$_{H,avr}$ is instead obtained from the reflection component, which can be modelled to derive the average properties of the obscuring medium, such as the above mentioned average torus column density, the obscuring medium covering factor ($C_{Tor}$) and its inclination angle ($\theta_{Inc}$) with respect to the observer. 

This paper is organised as follows: in Section \ref{sec:data}, we discuss the selection methods and data analysis techniques. In Section \ref{sec:model}, we present physically motivated torus models used in this work. Then, in Section \ref{sec:results}, we show the results of each of the sources we have analysed. In Section \ref{sec:Discussion_Conclusion}, we analyse and discuss our and display the current census of CT-AGN at $z<0.05$, combining our results with those obtained in previous works. Finally, in section \ref{sec:conclusions_summary}, we present the conclusions and a brief summary of our work, along with a mention about the future projects. In Appendix \ref{appendix1:tables} and \ref{appendix2:figures}, we show the tables of best-fit parameters and X-ray spectral fitting plots, respectively. All reported error ranges are at the $90\%$ confidence level, if not mentioned otherwise. Through the rest of the work, we assume a flat $\Lambda$CDM cosmology with H$_0$ = 69.6\,km\,s$^{-1}$\,Mpc$^{-1}$, $\Omega_m$=0.29 and $\Omega_\Lambda$=0.71 \citep{bennett14}. 

\section{Sample Selection and Data Reduction}\label{sec:data}

The seven sources (see Table \ref{tabel:1} for details) of our sample are CT-AGN candidates selected from the volume limited sample of Swift-BAT 100 month catalog, in the local Universe (z$<$ 0.05, $D\lesssim200$ Mpc). These Seyfert galaxies were previously classified as CT-AGN in \cite{Ricci_15_CTAGN-candidates} using the \bntorus\ model \citep{brightman2011xmm}, where \swx\ was used for E$<10$ keV (except for ESO138−G001, where \XMM\ data were used) and \swb\ for E$>10$ keV. Instead of BAT observations (15-150 keV), we are using \NuSTAR\ in the 3-50 keV range, as it is a grazing incidence telescope with lower background and smaller field of view resulting excellent sensitivity to source detection with better photon statistics. At E$<10$ keV, we preferred \XMM\ or \cha\ for better data quality, particularly in terms of source statistics compared to \swx\/. For NGC 2788A, only \swx\ data was available. The objects analysed in this work are CT-AGN candidates in the 100-month BAT sample for which no analysis with \MYTorus\ and \borus\ of the joint soft X-ray and NuSTAR spectra have been published yet.


\begin{sidewaystable*}
\center
\caption{Observational details of each sources.}
\label{tabel:1}
\vspace{.1cm}
  \begin{tabular}{ccccccccc}
       \hline
       \hline
    Source&AGN\tablefootmark{a}&log flux\tablefootmark{b}&Redshift&Instrument& Sequence& Start Time& Exposure Time& Net Spectral Counts\tablefootmark{c}\\ 
    &Types&in erg cm$^{-2}$ s$^{-1}$& (z) & & ObsID&(UTC)& (ks) & \\
    \hline
    &&&&\XMM& 0762920601& 2016-03-01& 30.0& 753, 225, 270\\
    \bf{MCG-02-12-017}&Sy2 candidate&8.84&0.03246&\NuSTAR& 60001160002& 2014-11-28& 34.0& 441, 379\\
    &&&&\NuSTAR& 60101015002& 2016-03-02& 19.5& 349, 369\\
       \hline
    &&&&\cha& 9438& 2008-11-16& 2.1& 6\\
    \bf{NGC 4180}&Sy2&17.58&0.00699&\NuSTAR& 60201038002& 2016-07-14& 23.4& 429, 387\\
    &&&&\NuSTAR& 60160480002& 2020-07-14& 31.6& 212, 212\\
       \hline  
    &&&&\swx&$*$& 2008---2020& 17.8\tablefootmark{d}& 34\\
    \bf{NGC 2788A}&Sy2&21.46&0.01335&\NuSTAR& 60469001002& 2019-06-14& 27.6& 608, 617\\
    &&&&\NuSTAR& 60160344002& 2020-08-14& 23.2& 639, 530\\   
       \hline
    &&&&\XMM& 0821870301& 2019-03-02& 32.6& 1129, 257, 422\\
    \bf{NGC 1106}&Sy2&15.49&0.01447&\NuSTAR& 60469002002& 2019-02-22& 18.7& 285, 332\\
    &&&&\NuSTAR& 60160130002& 2020-09-09& 22.3& 360, 354\\
       \hline 
    &&&&\cha& 14050& 2012-06-07& 5.1& 25\\
    \bf{ESO406-G004}&Sy2&12.38&0.02897&\NuSTAR& 60201039002& 2016-05-25& 36.3& 390, 357\\
    &&&&\NuSTAR& 60161799002& 2020-06-26& 23.7& 120, 86\\
       \hline
    &&&&\XMM& 0802450501& 2017-11-18& 44.8& 5232, 1980, 1134\\   
    \bf{2MASX}&Sy2&10.55&0.04422&\cha& 21299& 2018-12-17& 3.7& 532\\
    \bf{J20145928+2523010}&&&&\NuSTAR& 60201032002& 2017-05-27& 28.1& 1252, 1137\\
    &&&&\NuSTAR& 60160731002& 2020-04-21& 9.4& 608, 624\\
       \hline 
    &&&&\XMM& 0690580101& 2013-02-24& 135.4& 27058, 8878, 9016\\
    \bf{ESO138-G001}&Sy2&19.46&0.00914&\NuSTAR& 60201040002& 2016-05-22& 45.7& 3101, 2806\\
    &&&&\NuSTAR& 60061274002& 2020-04-01& 53.2& 3463, 3328\\
       \hline   
\end{tabular}
\par
\vspace{.2cm}
\tablefoottext{a,b}{AGN types and log flux between 14 and 195 keV is reported from the 105 month BAT catalog of \cite{Oh2018}. Only for ESO406-G004, the AGN type is reported from \cite{koss2016new}.}\\
\tablefoottext{c}{The reported \XMM\ net counts (background-subtracted total source counts) are those of the PN, MOS1 and MOS2 modules for a radius of 30\arcsec\ in 0.3--10\,keV, respectively. The reported \NuSTAR\ net counts are those of the FPMA and FPMB modules for a radius of 30\arcsec\ between 3--50\,keV, respectively. The reported \cha\ net counts are for the ACIS-I detector for a radius of 5\arcsec\ in 0.5--7\,keV. The reported \swx\ net counts are for a radius of 12\arcsec\ in 0.5--10\,keV.}\\
\tablefoottext{$*$}{Collection of the following ObsID- 00081038001, 00037312004, 00037312001, 00037312002, 00037312003, 03106140005, 03106140004, 03106140001, 03106140002, 07002346001, 07002347001.}\\
\tablefoottext{d}{Total exposure time of all the \swx\ observations}\\
\end{sidewaystable*}


\subsection{\NuSTAR\ Data Reduction}

We have used both focal plane modules FPMA, FPMB of \NuSTAR\ for each source. The collected data have been processed by NuSTAR Data Analysis Software-- NUSTARDAS version 2.0.0. The raw event files are calibrated by the \textit{nupipeline} script, using the response file from the Calibration Database-- CALDB version 20210202. The source and background spectra are extracted from a $30''$ ($\approx50\%$ of the encircled energy fraction--EEF at 10 keV) and $50''$ circular region respectively. Using \textit{nuproducts} scripts, we have generated source and background spectra files, along with ARF and RMF. Finally, the \NuSTAR\ spectra are grouped with at least 20 counts per bin, using \textit{grppha}. For each source, we have used all the available \NuSTAR\ observational data taken during different epochs, to a) check variability, and b) to improve the statistic of the spectra of these obscured sources between 3 and 50 keV.    

\subsection{\XMM\ Data Reduction} 

In \XMM\/, we have collected the data from the PN, MOS1 and MOS2 detectors. Using SAS version 19.0.0, we have processed the data using \textit{epproc} and \textit{emproc} for the PN and MOS filters, respectively. Finally we reduced and cleaned the flares using \textit{evselect}. The source photons were obtained from a $30''$ circular region, with $\sim85\%$ EEF for EPIC-PN at 1.5 keV. Background spectra were extracted from a $50''$ circle near the source. Each spectrum has been binned at 20 counts per bin, using \textit{grppha}. We prefer to use \XMM\ wherever it is available, because the effective area of \XMM\ in 0.3-10 keV is $\sim10$ times bigger than the \swx\ one and $\sim2$ times bigger than the \cha\/ one.

\subsection{\cha\ Data Reduction}

Although the effective area of \cha\ is smaller than that of \XMM\/, it is still ~5 times larger than \swx\/. Also, it has better angular resolution, a lower background and the higher capability of resolving extended emission from non-nuclear sources. We use \cha\ in two different scenarios: 1) when \XMM\ data are unavailable, and 2) when \XMM\ data are available, but to improve the photon statistics at E$<10$ keV. CIAO version 4.13 is used to process and reduce the data. The source spectra are extracted using a $5''$-radius circular region, that include $>99\%$ EEF. 

%
%
\section{Spectral Modeling}\label{sec:model}

 For X-ray spectral fitting on the objects in our sample, we have used \XSPEC\ \citep{arnaud1996xspec} version 12.11.1 in HEASOFT. The metal abundance is fixed to Solar metallicity from \cite{Anders_Grevesse1989}, while the photoelectric cross sections for all absorption components are obtained by using the approach of \cite{Verner1996}. The Galactic absorption column density is fixed for each source in our sample following \cite{Kalberla2005Gal}. We also used a thermal \mekal \citep{mewe1985calculated,Kaastra1992,Liedahl1995} component to phenomenologically model the soft excess which is often observed in the spectra of obscured AGN. 

We followed a standard approach in analysing the CT-AGN candidates, using self-consistent and up-to-date physically motivated uniform torus models, based on Monte Carlo simulations: \MYTorus\ \citep{murphy2009x,yaqoob2012nature} and \borus\ \citep{balokovic2018new}, which are specifically developed to characterise the X-ray spectra of heavily obscured AGN. In this Section, we describe how these two uniform torus models are used.

\subsection{MyTorus}\label{section:MYTorus}


The obscuring material in \MYTorus\ follows a toroidal or donut-like geometry, with circular cross-section. This model consists of three components: direct continuum (\textit{MYTZ}), Compton-scattered continuum (\textit{MYTS}) and fluorescent line component (\textit{MYTL}). The \textit{MYTZ}, also called zeroth-order component, models the attenuation of intrinsic X-ray radiation by the obscuring torus on the observer line-of-sight. The second component \textit{MYTS} computes the Compton-scattered photons, which are responsible for the ``Compton hump'' near $\sim20-30$ keV. Finally, \textit{MYTL} models prominent fluorescent emission lines such as: Fe K$_\alpha$ and Fe K$_\beta$ around 6.4 keV and 7.06 keV, respectively. Following the techniques in \cite{yaqoob2012nature} and from the previous results of \cite{marchesi2018compton,marchesi2019Nustar3,Marchesi2019Nustar5,Zhao2019Nustar2,Zhao2019Nustar4,Traina2021,Torres2021_CT-fraction,Silver2022}; we have used only the decoupled-configuration of \MYTorus\/, to estimate the clumpiness of the torus clouds. Here, we calculated the column density from direct continuum (N$\rm _{H,z}$) and scattered continuum (N$\rm _{H,S}$) separately, allowing flexibility on the parameter estimation even within a uniform cloud distribution framework.The ratio N$\rm _{H,z}$/N$\rm _{H,S}$ is used to evaluate the clumpiness, depending on how far the ratio is from unity. In \XSPEC\/, the configuration is as follows: 

\begin{equation}
\label{eq:MyT_edge}
\begin{aligned}
\mathrm{Model~{MyTorus_{edge-on}}} = C_{Ins}~*~phabs~*\\
(zpow * MYTZ + A_{S,90} * MYTS + A_{L,90} * MYTL + \\ 
f_s * zpow + mekal + zgauss),
\end{aligned}
\end{equation}

\begin{equation}
\label{eq:MyT_face}
\begin{aligned}
\mathrm{Model~{MyTorus_{face-on}}} = C_{Ins}~*~phabs~*\\
(zpow * MYTZ + A_{S,0} * MYTS + A_{L,0} * MYTL + \\ 
f_s * zpow + mekal + zgauss),
\end{aligned}
\end{equation}
\\

\noindent Here, equation \ref{eq:MyT_edge} models the Edge-On view ($\theta\rm _{Inc}=90\degree$) and equation \ref{eq:MyT_face} the Face-On view ($\theta\rm _{Inc}=0\degree$) on the AGN. We have used both inclination angles to compute a comparative study on the scattering column density arising from the polar dust (Edge-On) vs back-reflection of the torus (Face-On). We have equated and fixed the relative normalizations from scattering and line components, $A_{S}=A_{L}=1$, as we consider them to have been originated from the same regions where the direct power-law emerged. $C_{Ins}$ is a cross-calibration constant between the different instruments of telescopes (or a cross-normalization constant between different observations of same telescopes). We also included some additional components: $f_s$ to compute the scattering fraction from the direct powerlaw that does not interact (or elastically interact) with the torus, $mekal$ to phenomenologically model the soft excess, and $zgauss$ to include any additional emission lines.

\subsection{BORUS02}

The obscuring medium in \borus\ consists of a spherical geometry with bi-conical (polar) cut-out regions \citep{balokovic2018new}. This model is composed of three components: a) \borus\ itself which is a reprocessed component (including Compton-scattered $+$ fluorescent line component), b) $zphabs*cabs$ to include l.o.s. absorption with Compton scattering through the obscuring clouds. With this component we multiply a $cutoffpl_1$ to take into account the primary power-law continuum. C) Finally, another $cutoffpl_2$ component is included separately with $f_s$ to include a scattered unabsorbed continuum. The significant difference of \borus\ from \MYTorus\ is that the torus covering factor ($C _{Tor}$) in this model is kept as a free parameter varying from $0.1-1$ (i.e. the torus opening angle is ranging between $\theta\rm _{Tor}=0\degree-84\degree$), along with inclination angle $\theta\rm _{Inc}$ which is kept free between $18\degree-87\degree$. In our analysis using \XSPEC\/, we have used the following model configuration:  


\begin{equation}
\label{eq:borus02}
\begin{aligned}
\mathrm{Model~{borus02}} = C_{Ins} * phabs *( borus02 + zphabs\\
* cabs * cutoffpl_1 + f_s * cutoffpl_2\\
+ mekal + zgauss),
\end{aligned}
\end{equation}
\\

\noindent Here $mekal$ is included to compute the soft excess below $1$ keV, and $zgauss$ is introduced if there is any emission line signature not included in \borus\/.  

\section{Results of The X-ray Spectral Analysis}\label{sec:results}

In this Section, we are showing the results of X-ray spectral fitting on each CT-AGN candidate from \cite{Ricci_15_CTAGN-candidates}, using both physically motivated models mentioned in Section \ref{sec:model}, with two versions of \MYTorus\ and one \borus\/. In table \ref{Table:Results}, we display the summary of our analysis on the sample using \borus\/. The best fit parameters are reported in Table \ref{Table:best-fit_MCG-02-12-017} and in \ref{appendix1:tables}. The plots with X-ray spectral fitting are shown in figure \ref{fig:MCG-02-12-017} and \ref{appendix2:figures}. The background contribution for all these sources is within 20\%, unless mentioned otherwise. In the Tables, we also report the observed flux and intrinsic luminosity for each source. 


\begin{table}
\renewcommand*{\arraystretch}{1.5}
\centering
\caption{Summary of best-fit solutions of XMM-Newton and NuSTAR data using different models for MCG-02-12-017}
\label{Table:best-fit_MCG-02-12-017}
\vspace{.1cm}
   \begin{tabular}{ccccccc}
       \hline
       \hline       
       Model:&MyTorus&MyTorus&borus02\\
       &Edge-on&Face-on&&\\
    \hline
       $\chi^2$/dof&192/200&192/200&186/198\\
       $C_{Ins_1}$\tablefootmark{a}
       &0.88$_{-0.10}^{+0.11}$&0.89$_{-0.08}^{+0.09}$&0.89$_{-0.10}^{+0.11}$\\
       $C_{Ins_2}$\tablefootmark{b}
       &1.24$_{-0.15}^{+0.16}$&1.28$_{-0.14}^{+0.12}$&1.25$_{-0.15}^{+0.17}$\\
       $\Gamma$&1.94$_{-0.14}^{+0.14}$&1.98$_{-0.12}^{+0.06}$&2.11$_{-0.16}^{+0.13}$\\
       $C _{Tor}$\tablefootmark{c} &---&---&1.00$_{-0.35}^{+*}$\\
       $\theta\rm _{Inc}$\tablefootmark{d} &---&---&49$_{-*}^{+*}$\\
       N$\rm _{H,z}$\tablefootmark{e} &0.26$_{-0.03}^{+0.03}$&0.26$_{-0.02}^{+0.03}$&0.27$_{-0.03}^{+0.03}$\\
       N$\rm _{H,S}$\tablefootmark{f} &2.00$_{-1.10}^{+2.83}$&10.0$_{-9.99}^{+*}$&1.98$_{-0.52}^{+1.07}$\\
       $f_s$\tablefootmark{g} 10$^{-2}$&0.25$_{-0.13}^{+0.16}$&0.17$_{-0.09}^{+0.09}$&0.20$_{-0.11}^{+0.12}$\\
       kT\tablefootmark{h}&0.46$_{-0.37}^{+0.41}$&0.48$_{-0.19}^{+0.29}$&0.47$_{-*}^{+0.49}$\\
       F\tablefootmark{i}$_{2-10{\rm keV}}$ &5.65$_{-0.67}^{+0.40}$&5.59$_{-0.71}^{+0.44}$&5.55$_{-1.76}^{+0.45}$\\
       F\tablefootmark{j}$_{10-40{\rm keV}}$ &1.57$_{-0.36}^{+0.11}$&1.65$_{-0.41}^{+0.34}$&1.63$_{-0.60}^{+0.17}$\\
       L\tablefootmark{k}$_{2-10{\rm keV}}$ &4.98$_{-1.52}^{+2.13}$&4.82$_{-0.59}^{+0.56}$&5.17$_{-1.47}^{+1.89}$\\
       L\tablefootmark{l}$_{10-40{\rm keV}}$ &4.69$_{-1.44}^{+2.01}$&4.27$_{-0.51}^{+0.51}$&3.77$_{-1.07}^{+1.37 }$\\
       \hline
	\hline
	\vspace{0.02cm}
\end{tabular}

\tablefoot{We summarise here the best-fits of joint \XMM--\NuSTAR\ spectra using different torus models at 0.6-50 keV. The statistics and degrees of freedom for each fit are also reported.}

\tablefoottext{a}{$C_{Ins_1}$ = $C_{FPMA/PN}$ is the cross calibration constant between \NuSTAR\ observation of 2014 and \XMM\ observation of 2016.}\\
\tablefoottext{b}{$C_{Ins_2}$ = $C_{FPMA/PN}$ is the cross calibration constant between \NuSTAR\ observation of 2016 and \XMM\ observation of 2016.}\\
\tablefoottext{c}{Covering Factor: Fraction of sky covered by the torus, as seen by the nucleus, given by $C_{Tor}$ = cos($\theta_{\rm Tor}$).}\\
\tablefoottext{d}{Inclination Angle: The angle (in degrees) between symmetry axis of torus and line of sight angle}\\
\tablefoottext{e}{``Line of sight'' column density in $10^{24}$\,cm$^{-2}$.}\\
\tablefoottext{f}{Average column density from scattering in $10^{24}$\,cm$^{-2}$.}\\
\tablefoottext{g}{Fraction of primary emission getting scattered, rather than absorbed by the obscuring material.}\\
\tablefoottext{h}{Temperature in the thermal component \textit{mekal} in keV.}\\
\tablefoottext{i}{Flux between 2--10\,keV in $10^{-13}$ erg cm$^{-2}$ s$^{-1}$.}\\
\tablefoottext{j}{Flux between 10--40\,keV in $10^{-12}$ erg cm$^{-2}$ s$^{-1}$.}\\
\tablefoottext{k}{Intrinsic luminosity between 2--10\,keV in $10^{42}$ erg s$^{-1}$.}\\
\tablefoottext{l}{Intrinsic luminosity between 10--40\,keV in $10^{42}$ erg s$^{-1}$.}\\

\end{table}

\subsection{MCG-02-12-017}
\label{sec:MCG-02-12-017}

The source was marked as a CT-AGN candidate based on the data of \swx\ and \swb\/, with log N$_{\rm H,l.o.s.}$=24.25$_{-0.46}^{+1.06}$ in cm$^{-2}$. For our analysis, we have used the quasi-simultaneous observations of \XMM\ and \NuSTAR\/, along with another \NuSTAR\ observation which was taken about 15 months earlier with a longer exposure time $\sim34$ ks. The cross-calibration ratio between \XMM\/ and \NuSTAR\/ detector for quasi-simultaneous observations is $\sim75\%$ whereas for the previous NuSTAR observation, it is $\sim85\%$. 

\begin{figure*}
\begin{center}
\includegraphics[scale = 1.0, width = 18 cm, trim = 10 520 10 0, clip]{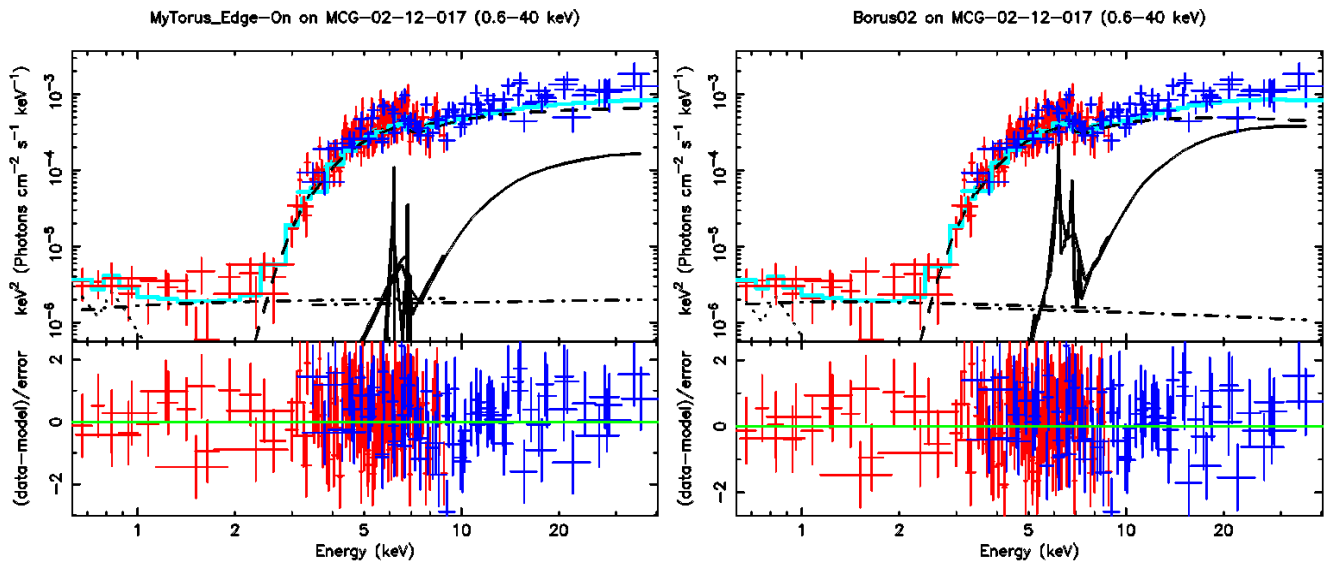} 
\caption{X-ray spectral fitting of decoupled edge-on \MYTorus\ (left) and \borus\ (right) models of MCG-02-12-017 data. In both the plots, the soft X-ray data (from \XMM\/) is marked in red and hard X-ray data (from \NuSTAR\/) is marked in blue. The joint best-fit model in both soft and hard X-rays is plotted as a cyan line. The individual model components are shown in black as follows: direct power-law emission (dash), reflected emission (solid), scattered emission (dot-dash), iron line (solid; in \MYTorus\/ it is separate, in \borus\ it is included in reflected emission) and mekal (dot).}
\label{fig:MCG-02-12-017}
\end{center}
\end{figure*}

This source is very well fitted (see Table \ref{Table:best-fit_MCG-02-12-017} and Figure \ref{fig:MCG-02-12-017}) by all three models. All the physically motivated models are in agreement that the observed line-of-sight column density N$\rm _{H,l.o.s.}=(0.23-0.30) \times 10^{24}$ cm$^{-2}$ is Compton-Thin, in disagreement with the \cite{Ricci_15_CTAGN-candidates} result. Even when we only used the quasi-simultaneous observations, the line-of-sight (l.o.s.) column density is consistent as N$\rm _{H,l.o.s.,qs}=(0.24-0.32) \times 10^{24}$ cm$^{-2}$. The average torus column density is instead found to be close or above the Compton thick threshold by \MYTorus\ Edge-On and \borus\ (N$\rm _{H,tor}=(0.9-4.83) \times 10^{24}$ cm$^{-2}$). The best-fit value of the photon index ranges between $\Gamma$=1.94-2.11, considering all the models. The estimation of torus properties, such as covering factor and opening angle in \borus\/, is found to be difficult since the reflection component is sub-dominant. 



\subsection{NGC 4180}
\label{sec:NGC4180}

 This target was classified as a CT-AGN candidate based on the data of \swx\ and \swb\/, log N$_{H}$=24.15$_{-0.22}^{+0.27}$ cm$^{-2}$. For our analysis, we included only the two \NuSTAR\ observations, excluding the \cha\ one due to its extremely poor photon statistic (Table \ref{tabel:1}). We used Portable Interactive Multi-Mission Simulator\footnote{\url{https://cxc.harvard.edu/toolkit/pimms.jsp}} to convert \NuSTAR\ spectrum for E$>$2 keV, and found the predicted count rate (1.56 $\times 10^{-3}$ cts/s) for \cha\ is within the error range of the observed one (1.42$\pm0.82 \times 10^{-3}$ cts/s). Moreover, the cross-normalization ratio between two separate observations (2016 and 2020) of the FPMA detector is $\sim 50\%$, portraying noticeable variability of the source.    

The source is very well fitted (see Table \ref{Table:best-fit_NGC4180} and Figure \ref{fig:NGC4180}) by all models, showing consistent results with each other, having Compton-Thick l.o.s. column density N$\rm _{H,l.o.s.}=(1.25-6.10) \times 10^{24}$ cm$^{-2}$ in agreement with the results obtained by \cite{Ricci_15_CTAGN-candidates}. Even the average torus column density, which is more accurately constrained by \MYTorus\ Edge-On and \borus\ in this case, shows N$\rm _{H,tor}=(0.66-4.56) \times 10^{24}$ cm$^{-2}$, suggesting a moderate CT nature of the obscuring material as a whole. The best-fit values of photon index are $\Gamma\sim 1.40-1.66$, considering all the models. The hard index value shows that the models might not be able to properly estimate the direct power-law component contribution, in absence of soft X-ray data, and therefore cannot fully break the N$\rm _{H,l.o.s.}$-$\Gamma$ degeneracy. For similar reasons, the covering factor and inclination angle also show a large range of uncertainty. However, as it is shown in Figure \ref{fig:NGC4180}, the overall spectral emission is dominated by the reflected component over the l.o.s. component. Therefore, from the available data, this source can be identified as a \textit{bona fide} CT-AGN. However, soft X-ray observations would be required to put stronger constraints on the different obscuring material parameters.

\subsection{NGC 2788A}
\label{sec:NGC2788A}

This source was marked as a CT-AGN candidate based on the data of \swx\ and \swb\/, with log N$_{H}$=25.55$_{-1.41}^{+*}$ in cm$^{-2}$. For our analysis, we have two \NuSTAR\ observations (taken in 2019 and 2020; total exposure $\sim 51$ ks). To cover the $<3$ keV energy range, we make use of 12 \swx\/ observations taken from 2008 to 2020\footnote{We obtained a joint spectrum using the tool available at \url{www.swift.ac.uk/user_objects/}}. Due to very low spectral counts in soft X-ray ($\sim34$ counts; see Table \ref{tabel:1}), we have groupped the spectra from XRT with 1 count/bin and jointly fitted the \swx\ and \NuSTAR\ spectra applying C-statistics over the entire range, in \XSPEC\/. The cross-calibration variability between \swx\ and \NuSTAR\ detectors fall within $\sim20\%$.

This source is very well fitted using the physically motivated models (see Table \ref{Table:best-fit_NGC2788A} and Figure \ref{fig:NGC2788A}). All models show consistent results between each other. The l.o.s. column density N$\rm _{H,l.o.s.}=(1.67-2.36) \times 10^{24}$ cm$^{-2}$ shows a CT column density, which validates the result of \cite{Ricci_15_CTAGN-candidates}, although with a significantly lower value. In comparison with the average column density of the torus, the best-fit value of \MYTorus\ Face-On and \borus\ are close to each other with the range N$\rm _{H,tor}=(1.72-22.69) \times 10^{24}$ cm$^{-2}$, agreeing with the CT nature of the cloud distribution. On these two models, the best-fit value of $\Gamma$ is around 1.8-1.9. In addition, the \borus\ model best fits the data with an intermediate covering factor, although with large uncertainties (0.49$_{-0.28}^{+0.47}$), and inclination angle in the range $47\degree-72\degree$. In Figure \ref{fig:NGC2788A}, both models show considerable dominance of the reflection component over the l.o.s. component, suggesting even more the CT nature of the source. From the available \NuSTAR\ data, this source is confirmed to be a \textit{bona fide} CT-AGN. However, we need more observations below 10 keV for a better understanding of the properties of the obscuring material.

\subsection{NGC 1106}
\label{sec:NGC1106}

This candidate was marked as a CT-AGN, based on the data of \swx\ and \swb\/, claiming it to have log N$_{H}$=24.25$_{-0.17}^{+0.29}$ in cm$^{-2}$. For our analysis, we have used \XMM\ observation (taken in 03/2019; $\sim33$ ks) and \NuSTAR\ observations (taken in 02-2019 and 09-2020; total exposure $\sim41$ ks). The cross-calibration ratio between \XMM\/ and \NuSTAR\/ detectors is $1.4$. 

This source is very well fitted (see Table \ref{Table:best-fit_NGC1106} and Figure \ref{fig:NGC1106}). All three models show consistent results with each other. The l.o.s. column density N$\rm _{H,l.o.s.}=(2.83-5.73) \times 10^{24}$ cm$^{-2}$ shows a CT column density, which validates the result of \cite{Ricci_15_CTAGN-candidates}. The torus average column density is N$\rm _{H,tor}=(1.34-7.98) \times 10^{24}$ cm$^{-2}$, in agreement with the CT nature of the torus. It is also interesting to note that \borus\/, which has better Reduced $\chi^2$ value ($\sim 1.04$) and a value of $\Gamma$ ($\sim 1.92$) closer to the AGN average \citep{Marchesi_Chandra2016}, also estimates that the l.o.s. column density is in agreement with the average torus column density, thus suggesting that the obscuring material is likely uniform (see Table \ref{Table:best-fit_NGC1106}). Besides, the \borus\ model gives a high covering factor (0.87$_{-0.24}^{+0.11}$), and moderate inclination angle between $28\degree-74\degree$. Figure \ref{fig:NGC1106} shows how the reflection component is dominant over the direct power-law. From the above analysis, this source can be counted as a \textit{bona fide} CT-AGN.

\subsection{ESO406-G004}
\label{sec:ESO406-G004}

This target was marked as a CT-AGN candidate based on the data of \swx\ and \swb\/, claiming to have log N$_{H}$=24.74$_{-0.55}^{+*}$ in cm$^{-2}$. For our analysis, we have a \cha\ observation with very low exposure ($\sim5.1$ ks) and only 25 spectral counts in soft X-rays. Furthermore, even though the two archival \NuSTAR\ observations have a much higher exposure time (total $\sim60$ ks), the source count statistic is significantly lower than that of the other sources ($\sim747$ net counts in 2016 and $\sim206$ net counts in 2020; see Table \ref{tabel:1}). Due to such low spectral counts, we have used C-Statistics to fit the data after binning with 1 count/bin in \cha\/, 10 counts/bin and 20 counts/bin on \NuSTAR\ observations of 2020 and 2016 respectively. It is also noticeable that the observations were taken after large gaps ($\sim 4$ years), and the cross-calibration ratio between ACIS (of \cha\/) and FMPA (of \NuSTAR\/) detectors show large variability $\sim1.3-1.7$. Furthermore, in the \NuSTAR\ observation of 06/2020, the background-noise contribution on the spectral signal is $\sim 30-40\%$, whereas the other data sets have $< 20\%$ contribution.

This source is very well fitted by all the models. All the three models show consistent results with each other. The l.o.s. column density for this source N$\rm _{H,l.o.s.}=(0.59-1.28) \times 10^{24}$ cm$^{-2}$ shows a mostly Compton-Thin column density, contrary to the result of \cite{Ricci_15_CTAGN-candidates}. The decoupled \MYTorus\ estimates N$\rm _{H,tor}=(0.13-2.03) \times 10^{24}$ cm$^{-2}$ and \borus\ estimates with large error range N$\rm _{H,tor}=(0.11-5.01) \times 10^{24}$ cm$^{-2}$. Overall, it shows mostly Compton-Thin clouds with upper bounds crossing the CT threshold. Due to the lack of \XMM\ data, it is likely that the low-statistic in the soft X-ray along with a fairly low statistic in the hard X-ray makes it difficult to properly disentangle the $\Gamma$-N$\rm _{H,l.o.s.}$ degeneracy. For similar reasons, we find that \borus\ computes low covering factor (best-fit value $\sim 0.10$), and inclination angle ($18\degree$) with high or unconstrained error range. 

Furthermore, we have noticed a significant cross-calibration variability for the two \NuSTAR\ observations (see Table \ref{Table:best-fit_ESO406-G004}). So, we also carried out a comparative study on the \NuSTAR\/ observations of this source taken in 05/2016 and 06/2020, to check flux and l.o.s. column density variability (following the approach of \citealt{Marchesi2022, Torres2023, Pizzetti2022}), in Table \ref{Table:flux_density_variability of ESO406-G004}. The cross-calibration flux value is measured with respect to the \cha\ observation of 06/2012. We fixed all the other parameters for these two \NuSTAR\ observations to the best-fit values, and only kept the N$_{H,l.o.s.}$ and flux free to vary. We studied the variability by fixing one of the two parameters and letting the other free to vary, and finally compared the values by varying both of them. We find that the N$_{H,l.o.s.}$ increases $\sim53\%$ from 2016 to 2020 and the flux is significantly increased ($\sim230\%$) in the 2020 observation with respect to the 2016 one. However, from the reduced C-stat value, we that find the residual (data-model) worsens if we vary only the l.o.s. column density. Whereas fixing the N$_{H,l.o.s.}$ does not significantly change the fit in respect to varying both flux and column density for the different \NuSTAR\ epochs. This indicates that the variability observed between the two \NuSTAR\ observations can be explained within a pure-luminosity variability scenario, while the fit improvement is not significant when allowing the l.o.s. column density free to vary.

\begin{table}
\renewcommand*{\arraystretch}{1.5}
\centering
\caption{Flux and column density variability of ESO406-G004 from \NuSTAR\ observations using \borus\/}
\label{Table:flux_density_variability of ESO406-G004}
\vspace{.01cm}
   \begin{tabular}{cccc}
       \hline
       \hline       
       Parameter&Fixing only&Fixing only&Varying both\\
       &N$_{H,l.o.s.}$&$C_{Ins}$&\\
    \hline
       $C_{Ins,NuS1}$\tablefootmark{a}&0.77$_{-0.12}^{+0.15}$&0.84$_{-0.16}^{+0.22}$&0.82$_{-0.14}^{+0.18}$\\
       $C_{Ins,NuS2}$\tablefootmark{b}&0.35$_{-0.07}^{+0.08}$&''&0.25$_{-0.08}^{+0.13}$\\
       N$_{H,l.o.s.,NuS1}$\tablefootmark{c}&0.82$_{-0.12}^{+0.15}$&0.88$_{-0.16}^{+0.21}$&0.86$_{-0.14}^{+0.17}$\\
       N$_{H,l.o.s.,NuS2}$\tablefootmark{d}&''&1.88$_{-0.45}^{+0.78}$&0.58$_{-0.21}^{+0.29}$\\
       \hline
       C-Stat/d.o.f.&$84/93$&$92/93$&$82/92$\\
	\hline
	\hline
	\vspace{0.01cm}
\end{tabular}
\tablefoottext{a}{Cross-calibration value from FPMA detector, observed in 2020}\\
\tablefoottext{b}{Cross-calibration value from FPMA detector, observed in 2016}\\
\tablefoottext{c}{l.o.s. column density from FPMA detector in $10^{24}$\/ cm$^{-2}$, , observed in 2020}\\
\tablefoottext{d}{l.o.s. column density from FPMA detector in $10^{24}$\/ cm$^{-2}$, observed in 2016}\\
\end{table}

\subsection{2MASX J20145928+2523010}
\label{sec:2MASXJ20145928+2523010}

This candidate was also classified as a CT-AGN, based on the data of \swx\ and \swb\/, claiming to have log N$_{H}$=24.42$_{-0.17}^{+0.20}$ in cm$^{-2}$. For our analysis, we have \cha\/ (taken in 12/2018) and \XMM\/ (taken in 11/2017) spectra with excellent photon statistics in the $0.6-10$keV energy range (total spectral counts $\sim 9$k). Even in hard X-rays, the two archival \NuSTAR\ observations (taken in 05/2017 and 04/2020) have a high exposure time and net spectral counts (total $\sim 3.6$k). It is worth noting, that for the joint \NuSTAR\ and \XMM\ observation taken in 2017, we measure a cross-calibration ratio $<1.4$. However, the flux of the 2018 \cha\ and the 2020 \NuSTAR\ observation is almost twice ($\sim1.93-2.15$) that of the 2017 \XMM\/ observation. There is also a significant flux variability (factor of 0.77) between the \NuSTAR\ and \XMM\ observations taken only 6 months apart.

\begin{table}
\renewcommand*{\arraystretch}{1.5}
\centering
\caption{Flux and column density variability of 2MASX J20145928+2523010 from \NuSTAR\ observations using \borus\/}
\label{Table:flux_density_variability of 2MASX J20145928+2523010}
\vspace{.01cm}
   \begin{tabular}{cccc}
       \hline
       \hline       
       Parameter&Fixing only&Fixing only&Varying both\\
       &N$_{H,l.o.s.}$&$C_{Ins}$&\\
    \hline
       $C_{Ins,NuS1}$\tablefootmark{a}&1.23$_{-0.04}^{+0.05}$&1.53$_{-0.07}^{+0.07}$&1.25$_{-0.05}^{+0.07}$\\
       $C_{Ins,NuS2}$\tablefootmark{b}&1.94$_{-0.09}^{+0.09}$&''&1.94$_{-0.09}^{+0.09}$\\
       N$_{H,l.o.s.,NuS1}$\tablefootmark{c}&*&5.63$_{-1.78}^{+1.96}$&0.33$_{-*}^{+1.61}$\\
       N$_{H,l.o.s.,NuS2}$\tablefootmark{d}&''&*&*\\
       \hline
       $\chi^2/d.o.f.$&$558/587$&$645/587$&$558/586$\\
	\hline
	\hline
	\vspace{0.01cm}
\end{tabular}
\tablefoottext{a}{Cross-calibration value from FPMA detector, observed in 2017}\\
\tablefoottext{b}{Cross-calibration value from FPMA detector, observed in 2020}\\
\tablefoottext{c}{l.o.s. column density from FPMA detector in $10^{22}$\/ cm$^{-2}$, observed in 2017}\\
\tablefoottext{d}{l.o.s. column density from FPMA detector in 2020 is unconstrained, represented as *}\\

\end{table}

This source is very well fitted by all the models (see Table \ref{Table:best-fit_2MASXJ20145928+2523010}). All the three models show consistent results with each other. The l.o.s. column density for this source is N$\rm _{H,l.o.s.}=(1.86-2.29) \times 10^{22}$ cm$^{-2}$, with a fairly low l.o.s. column density, just above the standard $10^{22}$ cm$^{-2}$ threshold which is used to classify obscured AGN. Such a result is in strong disagreement with the one by of \cite{Ricci_15_CTAGN-candidates}. Even the average column density of the torus is N$\rm _{H,tor}=(9.07-28.66) \times 10^{22}$ cm$^{-2}$, i.e. Compton-Thin. The decoupled \MYTorus\/ model shows a better estimate of the photon index $\sim 1.69-1.89$ compared to \borus\/, in terms of consistency with the expected value. Due to strong domination of the intrinsic powerlaw over the reflection component in the hard X-ray regime, \borus\ fails to compute the covering factor and inclination angle properly. Noticing the absence of a reflection component, we also tested a simple phenomenological model using photoelectric absorption and powerlaw above 3 keV. We found $\chi^2_{\nu}\sim0.99$ with N$_{H,l.o.s.}$ and $\Gamma$ within the error range of \borus\ results, considering a direct powerlaw along the line-of-sight. Thus, for the similarity of the results and to maintain consistency with the other sources, we have shown the results of physically motivated torus models only in Table \ref{Table:best-fit_2MASXJ20145928+2523010}.   

Furthermore, also for this source, we find a cross-calibration variability for different \NuSTAR\ observations. So, we show in Table \ref{Table:flux_density_variability of 2MASX J20145928+2523010} the comparative study on the \NuSTAR\/ observations of this source taken in 05/2017 and 04/2020, to check for flux and l.o.s. column density variability, similar to Table \ref{Table:flux_density_variability of ESO406-G004}. The cross-calibration value is measured with respect to the \XMM\ observation of 11/2017. Here also, we see the reduced $\chi^2$ value does not show any significant change when fixing only l.o.s. column density. But the $\chi^2/d.o.f.$ increases and worsens the fit, when we fix the cross-calibration parameter only. Therefore, similarly to the previous case, the observed flux change for this source also can be explained by the intrinsic luminosity variability. 

\subsection{ESO138-G001}
\label{sec:ESO138-G001}


\begin{table}[hbt!]
\renewcommand*{\arraystretch}{1.5}
\centering
\caption{Best-fitting parameters for different lines on data of ESO138-G001, for different torus models.}
\label{Table:Lines_ESO138-G001}
\vspace{.1cm}
   \begin{tabular}{ccccccc}
       \hline
       \hline       
       Lines&MyTorus&MyTorus&borus02\\
       &Edge-on&Face-on&&\\
    \hline
    \hline
       $EW$ of Mg XI &0.10$_{-0.01}^{+0.01}$&0.09$_{-0.01}^{+0.01}$&0.10$_{-0.01}^{+0.01}$\\
       Intensity of Mg XI 10$^{-5}$ &0.66$_{-0.06}^{+0.06}$&0.66$_{-0.06}^{+0.06}$&0.67$_{-0.06}^{+0.06}$\\
       \hline
       $EW$ of S XV &0.10$_{-0.02}^{+0.02}$&0.10$_{-0.02}^{+0.02}$&0.09$_{-0.01}^{+0.02}$\\
       Intensity of S XV 10$^{-5}$ &0.23$_{-0.05}^{+0.05}$&0.23$_{-0.05}^{+0.05}$&0.22$_{-0.05}^{+0.05}$\\
       \hline
       $EW$ of Si XIII &0.07$_{-0.01}^{+0.02}$&0.07$_{-0.01}^{+0.01}$&0.07$_{-0.01}^{+0.01}$\\
       Intensity of Si XIII 10$^{-5}$ &0.28$_{-0.04}^{+0.04}$&0.28$_{-0.04}^{+0.04}$&0.28$_{-0.04}^{+0.04}$\\

       \hline
	\hline
	\vspace{0.02cm}
\end{tabular}

\tablefoot{(1) We summarise here the details of the 3 most prominent emission lines in Figure \ref{fig:ESO138-G001} of joint \XMM--\NuSTAR\ spectra following the publications- \cite{De_Cicco2015,piconcelli2011x}.\\
(2) Equivalent Width ($EW$) of the lines are shown in keV. Normalization of line components are shown in photons/cm$^{-2}$\,s$^{-1}$.}

\end{table}

This source was marked as a CT-AGN based on the data of \XMM\ and \swb\/, claiming to have log N$_{H}$=25.25$_{-0.31}^{+*}$ cm$^{-2}$. For our analysis, we have used \XMM\ and \NuSTAR\ observations, both having excellent count statistic ($\sim45k$ counts in the $0.5-10$ keV and $\sim12.7k$ counts in the $3-50$ keV band, respectively). The cross-calibration ratio of the \NuSTAR\/ detector on \XMM\/ for the source is $\sim1.15$. 

The fit is worse (reduced $\chi^2\sim 1.33-1.45$; see Table \ref{Table:best-fit_ESO138-G001} and Figure \ref{fig:ESO138-G001}) than those measured in the other sources. The models are better fitted in the soft X-rays when adding all the emission lines listed in Table \ref{Table:Lines_ESO138-G001}, following the previous works of \cite{Piconcelli2011,DeCicco2015}. All three models are almost consistent with each other.

Studying all the models, the l.o.s. column density N$\rm _{H,l.o.s.}=(0.30-0.40) \times 10^{24}$ cm$^{-2}$ shows Compton-Thin clouds, differing from the results of \cite{Ricci_15_CTAGN-candidates}. In comparison with the average column density of torus, the decoupled \MYTorus\ (Face-On) and \borus\ estimates N$\rm _{H,tor}=(2.45-10.43) \times 10^{24}$ cm$^{-2}$, supporting a CT average column density scenario. It is also noticeable that the \borus\/ gives comparatively better Reduced $\chi^2$ value ($\sim 1.33$) and  $\Gamma\sim 1.95-1.99$. Besides, the \borus\ model further computes moderate-to-high covering factor ($0.68-0.83$), but low inclination angle with unconstrained error. We also had to include an extra Gaussian line profile for the fluorescent lines in the models, to account for a broader line profile than the one implemented within the torus models. The high N$\rm _{H,tor}$ and high covering factor show that the reprocessed emission is significantly dominant with prominent Fe line. In \borus\/, adding the Gaussian line improves the fit significantly, with F-statistic value$\approx42.2$.


\section{Discussions}
\label{sec:Discussion_Conclusion}

This paper reports the analysis of 7 CT-AGN candidates: MCG-02-12-017, NGC 2788A, NGC 4180, 2MASX J20145928+2523010, ESO406 G-004, NGC 1106 and ESO138 G-001 from the 100-month Palermo BAT sample. For the first time, we analysed the \NuSTAR\ spectra of these sources using \MYTorus\ and \borus\/.

\subsection{Clumpy Torus and Variability}


\begin{figure*}
\begin{center}
\includegraphics[scale = 1.0, width = 18 cm, trim = 50 270 50 270, clip]{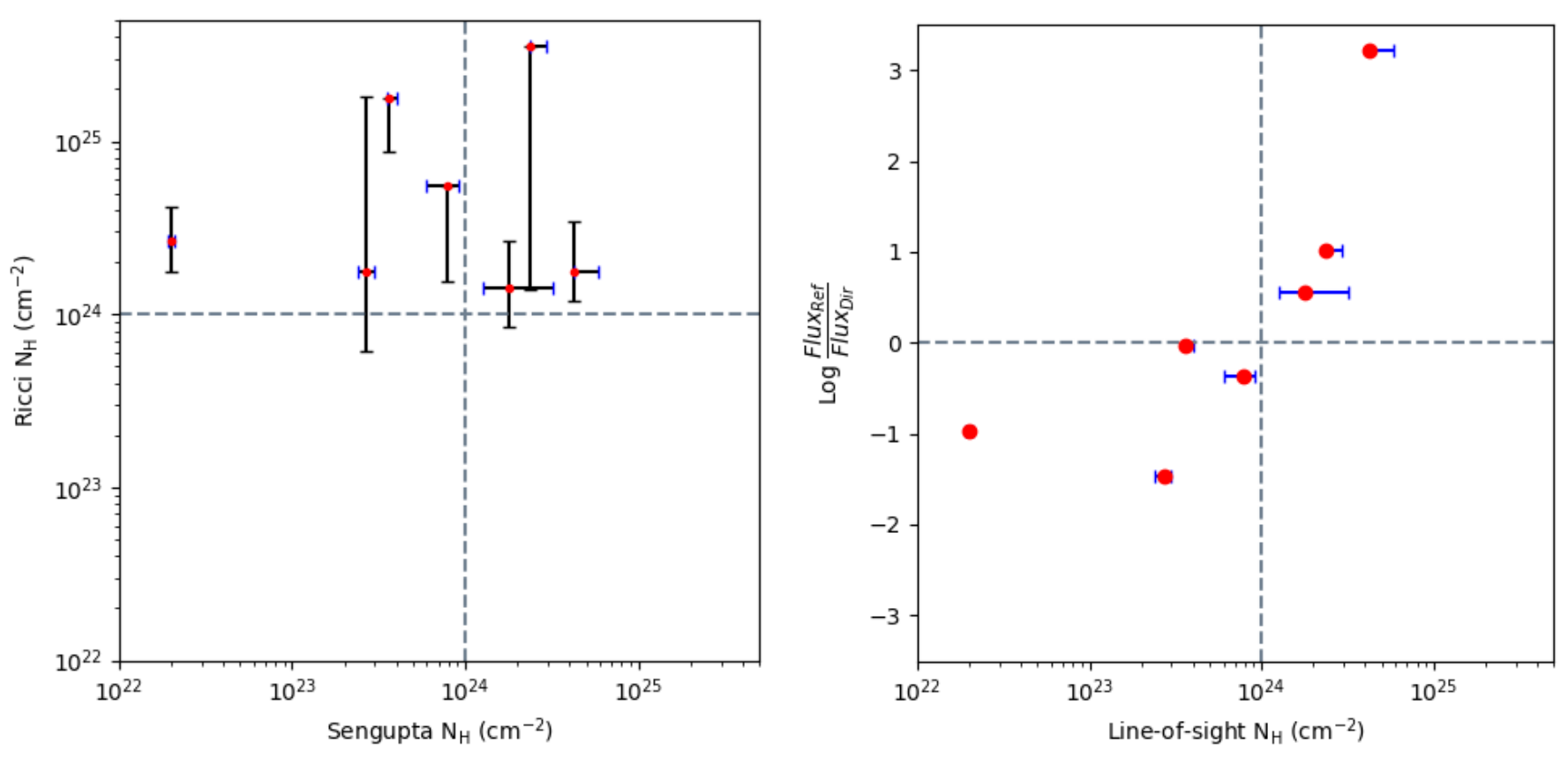} 
\caption{Comparisons of X-ray spectral properties of the seven sources. \textit{Left:} Comparing the l.o.s. column density values (as red-dot) and it's uncertainity values, from \cite{ricci2015compton} (black markers) versus this work (blue markers). The horizontal and vertical grey-dashed lines classify the CT column density threshold. \textit{Right}: Observed (i.e., non-absorption corrected) flux ratio of reflected component over the direct transmitted component in the 2-10 keV band for each source of our sample, plotted along the l.o.s. column density in X-axis. }
\label{fig:Ref_domination_and_Ricci_vs_Sengupta}
\end{center}
\end{figure*}

The l.o.s. column density and average torus column density of three out of seven sources in our sample (NGC 4180, NGC 2788A and NGC 1106) are found to be in agreement within their uncertainty ranges and above the Compton-Thick threshold ($>10^{24}$ cm$^{-2}$). For ESO406 G-004, the column densities are compatible within their error range, but fall in the Compton-Thin range. The remaining three sources (MCG-02-12-017, 2MASX J20145928+2523010, ESO138 G-001) show incompatible column densities thus hinting at a clumpy nature of the obscuring medium. In the left panel of Figure \ref{fig:Ref_domination_and_Ricci_vs_Sengupta}, we compare the l.o.s. column density results of our sample using \borus\ along with the results of \cite{ricci2015compton}. All the seven candidates from Ricci's paper lie above the CT threshold with large uncertainties, whereas the use of \NuSTAR\ data instead of \swb\ reduces the error bar significantly, displaying only three sources above CT line, leading to the confirmation of only three sources as CT. On the right side panel of Figure \ref{fig:Ref_domination_and_Ricci_vs_Sengupta} shows a clear trend of the flux ratio at 2-10 keV as a function of N$_{H,l.o.s.}$: the larger the l.o.s. column density, the stronger the flux of reflected continuum over direct continuum. 

Furthermore, in Figure \ref{fig:Census_CT}, we show the distribution of N$\rm _{H,l.o.s.}$ and N$\rm _{H,avr}$ of our sample AGN along with all the previous results of the CT-AGN candidates analysed by the Clemson-INAF group. The $\frac{N\rm _{H,l.o.s.}}{N\rm _{H,avr}}$=1 line is shown as a brown-dashed line and, 1:2 and 2:1 ratios are shown as pink dot-dashed lines to classify the sources with comparatively homogeneous torus when they produce the column density ratio within $\sim0.5-2.0$. Only 12 sources ($\sim22\%$ of the sample) fall within this region. The remaining 43 sources ($\sim78\%$ of the sample) show instead a significant inhomogeneity, and considering error bar at 3 $\sigma$ level, 34 sources ($\sim62\%$) fall completely outside the given area. This is also a natural outcome of a CCA scenario, in which the multi-phase clouds continuously rain through the meso scale, thus recurrently obscuring the line-of-sight. The residual gas experiencing less inelastic collisions (and thus less angular momentum cancellation) tends to accumulate in a clumpy torus-like structure at such meso scale (\citealt{Gaspari2017}). Therefore, from all the previous results along with those presented in this work, we can conclude that most of these obscured active galaxies have a significantly clumpy torus ($\sim78\%$ of the total population). It is also important to note that the two column densities are strongly non-correlated. By statistically analysing the parameters for all the 55 sources, their Pearson correlation coefficient\footnote{$\rho\approx 1$ or $\rho\approx -1$ for strong linear correlation or anti-correlation respectively, and $\rho\approx 0$ for lack of correlation.}  yields $\rho\approx 0.003$ (similar to what was obtained by \cite{Torres2021_CT-fraction}: -0.017). This states that a \textit{bona fide} CT-AGN should not necessarily be made of CT-torus. As supported by hydrodynamical simulations (see for example \citealt{Gaspari2020}), a realistic torus is a composition of multi-phase and multi-scale clouds, whose integral (e.g., density) can substantially change along each line of sight. This result of non-correlation is consistent even with the results of \cite{Zhao2021_NustarAGN}, in which a sample of $\sim 100$ local Compton-Thin AGN were studied (along with CT-AGN) using high quality \NuSTAR\ data along with soft X-ray data, finding that N$\rm _{H,tor}$ shows similar value ($\sim1.4\times10^{24}$ cm$^{-2}$) for different N$\rm _{H,l.o.s.}$. In Figure \ref{fig:Census_CT}, we increased our sample by including 74 sources from \cite{Zhao2021_NustarAGN}, marked as small grey circles. All of them have N$\rm _{H,l.o.s.}$ > 10$^{22}$ cm$^{-2}$. We found that the percentage of total population of homogeneous torus comes down to $\sim16\%$ of the enlarged sample, including the Compton-Thin sources. By calculating the Pearson correlation coefficient between the column densities with this enlarged sample, we find similar non-correlation scenario ($\sim-0.012$), as obtained before including these Compton-Thin AGN.

Through multi-epoch X-ray monitoring on these obscured sources, we can study the l.o.s. column density variability and confirm the in-homogeneity of the circum-nuclear cloud distribution. Some previous observations have reported extreme variability even like a ``changing-look'' nature from CT to Compton-Thin or vice-versa: NGC 7582 \citep{Bianchi2009_CT2CTn,Rivers2015_CT2CTn}, IC 751 \citep{Ricci2016_CT2CTn}, NGC 1358 \citep{Marchesi2022} and a few more. In our sample, 2MASX J20145928+2523010 shows strong variability over a three-year time-span. On the other hand, ESO138 G-001 shows almost no variability after 7 years of observational gap (Section \ref{sec:ESO138-G001}). We would need follow-up observations with longer exposures on all these sources, for a better study on variability (i.e. clumpiness) over the time scales from weeks to years. Nevertheless, to properly assess the complex cloud distribution within the torus, a joint analysis of both X-ray and mid-IR \citep{Berta2013,Buchner2019,Esparza-Arredondo2021} is required for an optimal study of each of these obscured sources along with the multi-epoch X-ray observations. We consider this for our future work.

\subsection{Updated Census of local CT-AGN candidates}


\begin{figure}
\centering
\includegraphics[width =\linewidth]{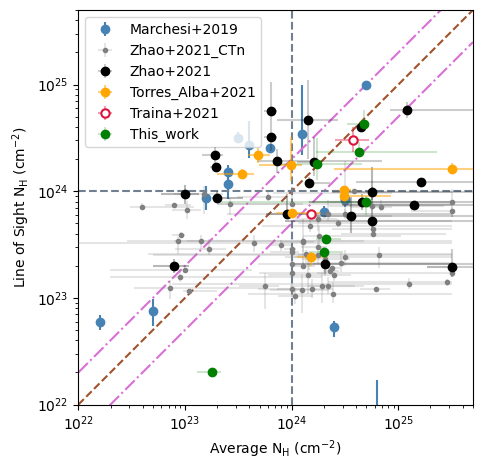} 
\caption{We show the census of the previous results of CT-AGN candidates (selected from \cite{ricci2015compton}) having z$<0.01$ with archival \NuSTAR\ data, analysed by the Clemson-INAF group: \cite{marchesi2018compton}, \cite{Zhao2021_NustarAGN}, \cite{Torres2021_CT-fraction}, \cite{Traina2021} and including our analysis. The sample of these CT-AGN candidates are marked as large circles. In the parameter space of average vs l.o.s. column density, grey-dashed lines drawn horizontally and vertically marks the CT column density threshold. The brown-dashed diagonal line (i.e the “Line of Homogeneity”) identifies an homogeneous obscuring material distribution. The region within the pink dot-dashed lines is used to classify the number of sources with homogeneous torus. We also included the sample of 74 Compton-Thin sources from \cite{Zhao2021_NustarAGN}, shown as small grey circles.}
\label{fig:Census_CT}

\end{figure}
 
 Out of the seven 100-month BAT candidate CT-AGN analysed in this work, three are confirmed to be \textit{bona fide} CT-AGN. This brings the total number of CT-AGN at $z<0.05$, to 35\footnote{\url{https://science.clemson.edu/ctagn/}} \citep{koss2016new, Oda2017, marchesi2018compton, marchesi2019Nustar3, Marchesi2019Nustar5, Georgantopoulos2019, tanimoto2019, Kammoun2020, Zhao2021_NustarAGN, Zhao2020_Xray_O3, Traina2021, Torres2021_CT-fraction}. In the left panel of Figure \ref{fig:Census_CT}, we studied 55 CT-AGN candidates analysed by our Clemson-INAF group in the parameter space of observed N$\rm _{H,l.o.s.}$ and computed N$\rm _{H,avr}$. 27 ($\sim50\%$) of them have N$\rm _{H,l.o.s.} > 10^{24}$ cm$^{-2}$.   
 
\begin{figure}
\centering
\includegraphics[width =\linewidth]{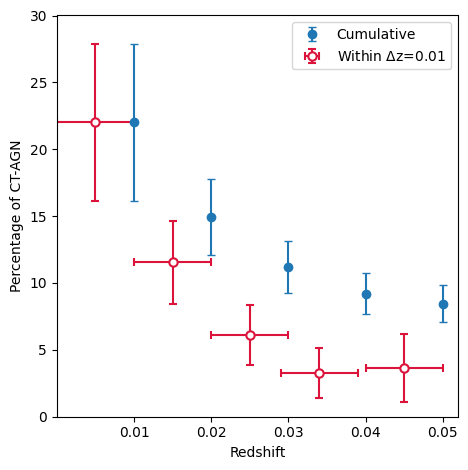}
\caption{Evolution of observed CT-AGN fraction from 100 month \swb\/ catalog as a function of redshift (for z$<0.05$). The red points represent the CT-AGN fraction within the given redshift bin of 0.01 and blue points show the cumulative value of the fraction within the given redshift. The displayed error bars are in Binomial statistics. This figure is updated from the CT-AGN fraction plot of \cite{Torres2021_CT-fraction}.}
\label{fig:Fraction_CT}
\end{figure}

The total percentage of confirmed CT-AGN from \swb\ selection within local Universe (z$<0.05$) is $\sim8\%$ (35/414), much lower than the CT-AGN fraction predicted by the population synthesis models. Our results also update the CT-AGN fraction within the distance $z<0.01$ to $(\sim22\%\pm5.9)\%$\footnote{Standard error in binomial distribution.} (11 CT-AGN out of 50 AGN; on \citealt{Torres2021_CT-fraction} the stats showed 10 CT-AGN out of 50). In figure \ref{fig:Fraction_CT}, we show the fraction of CT-AGN from the total AGN population in the 100 month \swb\/ catalog. The fraction drops by moving towards higher redshifts ($z>0.01$) because the CT-AGN sources become too faint to be detected by \swb\/ \citep{koss2016new}.


\subsection{Comparison with XClumpy results}

Recently, \citealt{Tanimoto2022} (T22 hereafter) published the results of the X-ray spectral analysis with the \textit{`XClumpy'} model \citep{Tanimoto2019_Xclumpy} of the sources analysed in this work (as part of a larger sample of low--redshift, heavily obscured AGN). \textit{`XClumpy'} model considers the torus as clumpy and in-homogeneous, assuming power-law distribution along the radial axis and Gaussian distribution along the vertical axis of the torus. Below, we compare their results with those obtained using \MYTorus\ decoupled and \borus\/ in this work.


\begin{table*}[hbt!]
\renewcommand*{\arraystretch}{1.5}
\begin{center}
\caption{Comparisons of best-fit values for $\Gamma$ (photon index) and Torus column density using the torus models}
\label{Table:borusVsMytorusVsXclumpy}
\vspace{.1cm}
    \begin{tabular}{ccc|cc|cccc}
       \hline
       \hline  
       Models:&\borus\/&&\MYTorus\/&(Decoupled Edge-On)&\textit{XClumpy}& (from T22)&\\
       \hline
       Sources&$\Gamma$&N$\rm _{H,torus}$&$\Gamma$&N$\rm _{H,S}$&$\Gamma$&N$\rm _{H,eq}$\\
       &&($10^{24}$ cm$^{-2}$)&&($10^{24}$ cm$^{-2}$)&&($10^{24}$ cm$^{-2}$)&\\
    \hline
    \hline
       MCG-02-12-017&2.11$_{-0.16}^{+0.13}$&1.98$_{-0.52}^{+1.07}$&1.94$_{-0.14}^{+0.14}$&2.00$_{-1.10}^{+2.83}$&1.54$_{-0.45}^{+0.18}$&9.59$_{-8.43}^{+*}$\\
       NGC4180&1.55$_{-*}^{+0.44}$&1.74$_{-1.08}^{+2.82}$&1.66$_{-*}^{+0.39}$&1.97$_{-0.73}^{+0.33}$&1.61$_{-0.17}^{+0.05}$&4.10$_{-2.18}^{+*}$\\
       NGC2788A&1.95$_{-0.31}^{+0.32}$&4.26$_{-2.54}^{+18.43}$&1.56$_{-*}^{+0.20}$&1.25$_{-0.60}^{+0.24}$&1.67$_{-0.08}^{+0.11}$&2.37$_{-0.19}^{+2.20}$\\
       NGC1106&1.92$_{-0.35}^{+0.44}$&4.83$_{-1.38}^{+*}$&1.40$_{-*}^{+0.00}$&1.46$_{-0.12}^{+0.18}$&1.67$_{-0.30}^{+0.16}$&8.17$_{-1.51}^{+1.63}$\\
       ESO406-G004&1.42$_{-*}^{+0.02}$&4.97$_{-4.80}^{+*}$&1.48$_{-*}^{+0.30}$&0.49$_{-0.62}^{+1.11}$&1.10$_{-*}^{+0.28}$&0.64$_{-0.23}^{+0.55}$\\
       2MASX J20145928+2523010&1.52$_{-0.04}^{+0.05}$&0.18$_{-0.04}^{+0.05}$&1.79$_{-0.10}^{+0.10}$&0.12$_{-0.03}^{+0.04}$&1.42$_{-0.09}^{+0.03}$&1.01$_{-0.67}^{+0.89}$\\
       ESO138-G001&1.98$_{-0.03}^{+0.01}$&10.35$_{-0.08}^{+0.08}$&1.53$_{-0.05}^{+0.05}$&1.47$_{-0.49}^{+0.60}$&1.45$_{-0.06}^{+0.03}$&5.34$_{-3.66}^{+1.53}$\\
	\hline
	\hline
	\vspace{0.1cm}
    \end{tabular}

\end{center}
\end{table*}

\begin{itemize}
    
    \item \textbf{MCG-02-12-017:} For this source, T22 used the quasi-simultaneous observations of \XMM\ and \NuSTAR\/ (ObsID: 60101015002). The l.o.s. column density and photon index, at $90\%$ confidence, gives N$\rm _{H,l.o.s.}=(0.21-0.28) \times 10^{24}$ cm$^{-2}$ and $\Gamma=1.53-1.88$, respectively. The results are very consistent with our results (see Table \ref{Table:best-fit_MCG-02-12-017}). Similarly, the computed equatorial (average) column density mostly falls within the limits of error range (N$\rm _{H,eq}=(1.16-9.59) \times 10^{24}$ cm$^{-2}$) of our analysis, which is also in agreement with our prediction of clumpy nature of the torus. Overall, for this source, the \textit{XClumpy} model is in agreement with the decoupled \MYTorus\ and \borus\ models.
    
    \item \textbf{NGC 4180:} For this source, T22 used the observations of \cha\ and \NuSTAR\/ (ObsID: 60201038002). The l.o.s. column density and photon index, at the $90\%$ confidence level, gives N$\rm _{H,l.o.s.}=(0.68-3.75) \times 10^{24}$ cm$^{-2}$ and $\Gamma=1.44-1.66$, respectively. The results are consistent with the decoupled Face-ON \MYTorus\ and \borus\ results (see Table \ref{Table:best-fit_NGC4180}). Similarly, the computed equatorial (average) column density mostly falls within the range (N$\rm _{H,eq}=(1.92-4.10) \times 10^{24}$ cm$^{-2}$) computed in our analysis, which is also in agreement with our prediction of clumpy nature of the torus. Even though in our analysis, we only used \NuSTAR\ data on this source, the results of \MYTorus\ and \borus\ model still comes in agreement with the \textit{XClumpy} results.
    
     \item \textbf{NGC 2788A:} For this source, T22 used the observations of \textit{Suzaku} (obsID: 710007010) and \NuSTAR\/ (ObsID: 60469001002). The l.o.s. column density and photon index, at the $90\%$ confidence level, gives N$\rm _{H,l.o.s.}=(1.55-2.83) \times 10^{24}$ cm$^{-2}$ and $\Gamma=1.59-1.78$, respectively. The results are very consistent with our results (see Table \ref{Table:best-fit_NGC2788A}), even though we have used \swx\ instead of \textit{Suzaku} at energies E$<10$ keV. The computed equatorial (average) column density also falls within the range (N$\rm _{H,eq}=(2.18-4.57) \times 10^{24}$ cm$^{-2}$) of our analysis. Even for the covering factor measurement, \textit{XClumpy} estimates $\theta\rm _{Tor}=19\degree-46\degree$, which significantly narrows the error range and falls within the \borus\ computed range of $\theta\rm _{Tor}=16\degree-78\degree$. Also, while computing the inclination angle $\theta\rm _{Inc}=62\degree-85\degree$, the values can be considered in agreement within the uncertainties with \borus\ fitting ($47\degree-72\degree$). Overall, for this source, the results obtained using the \textit{XClumpy} model can be regarded as consistent with those obtained using the decoupled \MYTorus\ and \borus\ ones.


\begin{table*}[hbt!]
\renewcommand*{\arraystretch}{1.5}
\centering
\caption{Best fit \borus\ parameters for the sample}
\label{Table:Results}
\vspace{.1cm}
   \begin{tabular}{ccccccc}
       \hline
       \hline  
       Sources&$\Gamma$&N$\rm _{H,l.o.s.}$&N$\rm _{H,torus}$&$C _{Tor}$&$\theta\rm _{Inc}$\\
       &&($10^{24}$ cm$^{-2}$)&($10^{24}$ cm$^{-2}$)&&(Degees)&\\
    \hline
    \hline
       MCG-02-12-017&2.11$_{-0.16}^{+0.13}$&0.27$_{-0.03}^{+0.03}$&1.98$_{-0.52}^{+1.07}$&1.00$_{-0.35}^{+*}$&49$_{-*}^{+*}$\\
       NGC4180&1.55$_{-*}^{+0.44}$&1.78$_{-0.51}^{+1.40}$&1.74$_{-1.08}^{+2.82}$&0.88$_{-0.77}^{+*}$&49$_{-*}^{+12}$\\
       NGC2788A&1.95$_{-0.31}^{+0.32}$&2.34$_{-0.58}^{+*}$&4.26$_{-2.54}^{+18.43}$&0.49$_{-0.28}^{+0.47}$&63$_{-16}^{+9}$\\
       NGC1106&1.92$_{-0.35}^{+0.44}$&4.79$_{-1.96}^{+*}$&4.83$_{-1.38}^{+*}$&0.87$_{-0.24}^{+0.11}$&37$_{-9}^{+37}$\\
       ESO406-G004&1.42$_{-*}^{+0.02}$&0.79$_{-0.12}^{+0.04}$&4.97$_{-4.80}^{+*}$&0.10$_{-*}^{+*}$&18$_{-*}^{+69}$\\
       2MASX J20145928+2523010&1.52$_{-0.04}^{+0.05}$&0.02$_{-0.00}^{+0.00}$&0.18$_{-0.05}^{+0.04}$&1.00$_{-0.23}^{+*}$&18$_{-*}^{+*}$\\
       ESO138-G001&1.98$_{-0.03}^{+0.01}$&0.47$_{-0.06}^{+0.07}$&10.35$_{-0.08}^{+0.08}$&0.80$_{-0.12}^{+0.03}$&18$_{-*}^{+*}$\\
	\hline
	\hline
	\vspace{0.1cm}
\end{tabular}

\end{table*}

    \item \textbf{NGC 1106:} For this source, T22 used the observations of \XMM\ and \NuSTAR\/ (ObsID: 60469002002). The l.o.s. column density and photon index, at $90\%$ confidence, shows N$\rm _{H,l.o.s.}=3.45-4.29 \times 10^{24}$ cm$^{-2}$ and $\Gamma=1.37-1.83$, respectively. These results are consistent with our results (see Table \ref{Table:best-fit_NGC1106}). On the other hand, the computed equatorial (average) column density falls in the upper limits of error range (N$\rm _{H,eq}=(6.66-9.80) \times 10^{24}$ cm$^{-2}$) compared to our analysis. In case of computing the inclination angle $\theta\rm _{Inc}=60\degree-77\degree$ and covering factor $\theta\rm _{Tor}=14\degree-31\degree$, their results falls within the large confidence range of our analysis ($\theta\rm _{Inc}=28\degree-74\degree$ and $\theta\rm _{Tor}=11\degree-51\degree$). It is noticeable that the error ranges are significantly reduced while using \textit{XClumpy}. Nevertheless, for this source, the \textit{XClumpy} model results are compatible with those obtained using the decoupled \MYTorus\ and \borus\ models.
    
    \item \textbf{ESO406 G-004:} For this source, T22 used the observations of \swx\ (ObsID: 00081420001) and \NuSTAR\/ (ObsID: 60161799002). The l.o.s. column density and photon index, at the $90\%$ confidence level, shows N$\rm _{H,l.o.s.}=(0.38-6.34) \times 10^{24}$ cm$^{-2}$ and $\Gamma=1.10-1.38$, respectively. As we used \cha\ and both available \NuSTAR\ data, our results show better constraints (see Table \ref{Table:best-fit_ESO406-G004}) on this source compared to their results. However, the computed equatorial (average) column density mostly falls within the range (N$\rm _{H,eq}=(0.41-1.19) \times 10^{24}$ cm$^{-2}$) of our analysis. The \textit{XClumpy} model is compatible with decoupled \MYTorus\ and \borus\ results.

    \item \textbf{2MASX J20145928+2523010:} For this source, T22 used the observations of \XMM\ and \NuSTAR\/ (ObsID: 60201032002). The l.o.s. column density and photon index, at $90\%$ confidence, gives N$\rm _{H,l.o.s.}=(0.01-0.03) \times 10^{24}$ cm$^{-2}$ and $\Gamma=1.33-1.45$, respectively. The results of N$\rm _{H,l.o.s.}$ shows consistency with our results, but $\Gamma$ value is much lower than \MYTorus\ and \borus\/ (see Table \ref{Table:best-fit_2MASXJ20145928+2523010}). However, the computed equatorial (average) column density falls within the range (N$\rm _{H,eq}=(0.34-1.90) \times 10^{24}$ cm$^{-2}$) of our analysis, which is also in agreement with our prediction of clumpy torus clouds. Overall, for this source, the \textit{XClumpy} model is consistent with most of the results of decoupled \MYTorus\ and \borus\ models, except in $\Gamma$.
    
    \item \textbf{ESO138 G-001:} For this source, T22 used the observations of \XMM\ and \NuSTAR\/ (ObsID: 60201040002). The l.o.s. column density and photon index, at $90\%$ confidence, gives N$\rm _{H,l.o.s.}=(0.42-0.50) \times 10^{24}$ cm$^{-2}$ and $\Gamma=1.42-1.51$, respectively. The results are quite inconsistent with our results (see Table \ref{Table:best-fit_ESO138-G001}), specially compared to the photon index from \borus\/ which is $\Gamma=1.95-1.99$ . Similarly, the computed equatorial (average) column density has a value within the range (N$\rm _{H,eq}=(1.68-6.87) \times 10^{24}$ cm$^{-2}$) of our analysis. Their covering factor $\theta\rm _{Tor}=10\degree-13\degree$ lies much below than the \borus\ estimates of $\theta\rm _{Tor}=34\degree-47\degree$. For this source, the \textit{XClumpy} model in T22 is inconsistent with the decoupled \MYTorus\ and \borus\ results. We note that, \borus\ results in a better fit (reduced $\chi^2$ value $\sim 1.32$) compared to the fit exhibited in T22 paper (reduced $\chi^2$=1.39).
    
\end{itemize}

Summarizing the above comparisons, we find \textit{XClumpy} places stronger constraints on the different torus parameters, by having smaller uncertainties, with respect to \borus\ and \MYTorus\/. The only exception is ESO138 G-001, which is best-fitted using \borus\/ in our analysis, because T22 did not include the prominent emission lines that we mentioned in Table \ref{Table:Lines_ESO138-G001}. We also noticed that \textit{XClumpy} model shifts the best-fit value of photon index to harder values by $\sim6-22\%$ compared to \borus\ and by  $\sim0.6-26\%$ compared to the best-fit model of decoupled \MYTorus\/, in our sample. In table \ref{Table:borusVsMytorusVsXclumpy}, we have shown the values of photon index and average torus column density, computed using different torus models for each source. We note that T22 used only one \NuSTAR\ observation for each source. In this work, we have used all the available observations from \NuSTAR\ for these heavily obscured sources, to increase the photon statistic over 10keV and minimise the under/over-estimation of spectral parameters. Overall, for our sources, we do not find any significant discrepancies when using \textit{XClumpy} in comparison with \borus\ and decoupled \MYTorus\ models.

\section{Conclusions and Summary}
\label{sec:conclusions_summary}

In this work, we studied and classified 7 CT-AGN candidates from 100 month \swb\/ catalog using archival \NuSTAR\/ observations. All sources have at least one \NuSTAR\ observation covering the 3--50\, keV energy range. In the 0.6--10\,keV band, we have used \XMM\ data for three targets, \cha\ data for two targets and both \XMM\/--\cha\ in one target. NGC 2788A has only \swx\ data in the soft X-ray. We classified the sources on the basis of their best-fit value of l.o.s. hydrogen column density, i.e., if N$\rm _{H,l.o.s.} \geq 10^{24}$ cm$^{-2}$ the candidates are marked as \textit{bona-fide} CT-AGN. Otherwise they are identified as partially CT-AGN or Compton-Thin AGN depending on their column density. The summary of our results and conclusions are as follows:
\begin{enumerate}

    \item From the 7 CT-AGN candidates, 3 are confirmed as \textit{bona fide} CT-AGN with moderate to high covering factors, using \NuSTAR\ data above 10 keV. Three of them shows l.o.s. column density Compton-Thin, but torus column density above CT-threshold. Only 2MASX J20145928+2523010 shows Compton-Thin values in both the column densities. The summary of the results of all these sources with \borus\ model is displayed in Table \ref{Table:Results}.
    
    \item This present work updates the total number of \NuSTAR\/-confirmed CT-AGN to 35, for $z<0.05$, which is $\sim8\%$ of the total AGN population in 100 month BAT catalog. This value is still quite below the predicted value ($\sim20-50\%$) of the CXB population synthesis models, which suggests that a significant fraction of heavily obscured AGN are missed even by a hard X-ray telescope such as \swb\/.
    
    \item Out of 55 CT-AGN candidates analysed by our Clemson-INAF research group, adding the results of this work brings the population of \textit{bona fide} CT-AGN to 27 ($\sim50\%$). Among these, only 14 ($\sim25\%$) candidates show both N$\rm _{H,l.o.s.}$ and N$\rm _{H,avr}$ above the CT-threshold.
    
    \item We find no correlation between these two column densities (N$\rm _{H,l.o.s.}$ and N$\rm _{H,avr}$) from our sample. Our results state that identifying a \textit{bona fide} CT-AGN i.e. obscured AGN with N$_{H,l.o.s.} >$ 10$^{24}$ cm$^{-2}$ would not lead to the conclusion that the torus is also CT. Similarly, Compton-Thin N$_{H,l.o.s.}$ should not mean that the torus is also Compton-Thin.
    
    \item Most of these obscured galaxies have significantly clumpy or in-homogeneous distribution of clouds. Multi-epoch monitoring of these sources using telescopes like \XMM\/, \cha\/ and \NuSTAR\/ will help us to study their intrinsic flux and l.o.s. column density variability. This will give a better understanding of cloud movements in the obscuring medium and X-ray emission from the central engine.
    
    \item \MYTorus\ and \borus\ results are consistent with each other, in estimating the column densities and other parameters of the sources. In most of the cases, \borus\ shows better fitting from a statistical point of view, along with the estimation of torus opening angle and inclination angle of the obscured AGN which are fixed parameters in \MYTorus\/.
    
    \item We find our results on the 7 CT-AGN candidates using the uniform torus models are compatible with the results of the non-uniform torus model \textit{XClumpy} in \cite{Tanimoto2022}. However we also notice the trend that \textit{XClumpy} shifts the photon index to harder values in comparison to the uniform torus models we have used. As proof, we have displayed the $\Gamma$ and torus column density values of T22 with respect to our results in table \ref{Table:borusVsMytorusVsXclumpy}.

\end{enumerate}
    
For future works, a joint analysis of the X-ray properties and the reprocessed emission at mid-IR will be carried out for a better understanding of the torus structure and obscuration properties (Sengupta et al. in prep.). We will also use multi-epoch observations on some of these sources, in the framework of theoretical models like CCA (\citealt{Gaspari2013}), warped-accretion disk (\citealt{Buchner2021}) and others, in the future works (\citealt{Torres2023}, Sengupta et al. in prep., Pizzetti et al. in prep).
    

\section{Acknowledgements}

We thank the referee for the useful suggestions that helped
us in improving the paper. This research has made use of the \NuSTAR\ Data Analysis Software (NuSTARDAS) jointly developed by the ASI Space Science Data Center (SSDC, Italy) and the California Institute of Technology (Caltech, USA). 
DS acknowledges the PhD and ``MARCO POLO UNIVERSITY PROGRAM'' funding from the Dipartimento di Fisica e Astronomia (DIFA), Università di Bologna. DS and SM acknowledge the funding from INAF ``Progetti di Ricerca di Rilevante Interesse Nazionale'' (PRIN), Bando 2019 (project: ``Piercing through the clouds: a multiwavelength study of obscured accretion in nearby supermassive black holes''). MG acknowledges partial support by HST GO-15890.020/023-A and the {\it BlackHoleWeather} program.

\bibliographystyle{aa}
\bibliography{1biblio}

\begin{appendix}
\section{Tables of X-Ray Spectra}

\label{appendix1:tables}

\begin{table}[hbt!]
\renewcommand*{\arraystretch}{1.5}
\centering
\caption{Summary of best-fit solutions of NuSTAR data using different models for NGC 4180}
\label{Table:best-fit_NGC4180}
\vspace{.1cm}
   \begin{tabular}{ccccccc}
       \hline
       \hline       
       Model&MyTorus&MyTorus&borus02\\
       &Edge-on&Face-on&&\\
    \hline
       $\chi^2$/dof&76/64&84/64&76/62\\
       $C_{Ins}$\tablefootmark{a}       &1.47$_{-0.21}^{+0.27}$&1.48$_{-0.22}^{+0.26}$&1.47$_{-0.21}^{+0.26}$\\
       $\Gamma$&1.66$_{-*}^{+0.39}$&1.40$_{-*}^{+0.22}$&1.55$_{-*}^{+0.44}$\\
       $C _{Tor}$&---&---&0.88$_{-0.77}^{+*}$\\
       $\theta\rm _{Inc}$ &---&---&49$_{-*}^{+12}$\\
       N$\rm _{H,z}$ &6.10$_{-4.30}^{+*}$&1.49$_{-0.24}^{+0.36}$&1.78$_{-0.51}^{+1.40}$\\
       N$\rm _{H,S}$ &1.97$_{-0.73}^{+0.33}$&3.98$_{-*}^{+1.08}$&1.74$_{-1.08}^{+2.82}$\\
       $f_s$10$^{-2}$ &0.40$_{-0.20}^{+1.02}$&0.55$_{-*}^{+1.05}$&2.27$_{-1.68}^{+1.31}$\\
       F$_{2-10{\rm keV}}$ &1.34$_{-*}^{+0.22}$&1.30$_{-*}^{+0.15}$&1.33$_{-*}^{+3.82}$\\
       F$_{10-50{\rm keV}}$ &4.49$_{-*}^{+20.85}$&4.73$_{-*}^{+9.22}$&4.54$_{-*}^{+0.39}$\\
       L$_{2-10{\rm keV}}$\tablefootmark{b} &11.01$_{-*}^{+51.43}$&2.72$_{-0.47}^{+3.18}$&3.21$_{-1.38}^{+7.78}$\\
       L$_{10-50{\rm keV}}$\tablefootmark{c} &19.05$_{-*}^{+89.05}$&7.14$_{-4.27}^{+0.38}$&6.63$_{-2.85}^{+15.99}$\\
       \hline
	\hline
	\vspace{0.02cm}
\end{tabular}

\tablefoot{We summarise here the best-fits of \NuSTAR\ spectra using different torus models between 3 and 50 keV, referred in Section \ref{sec:NGC4180}. The statistics and degrees of freedom for each fit are also reported. The parameters are reported as in Table \ref{Table:best-fit_MCG-02-12-017} if not mentioned otherwise.}\\
\tablefoottext{a}{$C_{Ins}$ is the ratio of cross-normalization constant between two \NuSTAR\ observations through their FPMA detectors.}\\
\tablefoottext{b}{Intrinsic luminosity between 2--10\,keV in $10^{41}$ erg s$^{-1}$.}\\
\tablefoottext{c}{Intrinsic luminosity between 10--50\,keV in $10^{41}$ erg s$^{-1}$.}\\

\end{table}

\begin{table}[hbt!]
\renewcommand*{\arraystretch}{1.5}
\centering
\caption{Summary of best-fit solutions of \swx\ and NuSTAR data using different models for NGC 2788A}
\label{Table:best-fit_NGC2788A}
\vspace{.1cm}
   \begin{tabular}{ccccccc}
       \hline
       \hline       
       Model&MyTorus&MyTorus&borus02\\
       &Edge-on&Face-on&&\\
    \hline
       C-Stat/dof&136/153&116/153&116/151\\
       $C_{Ins_1}$\tablefootmark{a} &1.12$_{-0.38}^{+0.58}$&0.93$_{-0.26}^{+0.40}$&0.94$_{-0.27}^{+0.40}$\\
       $C_{Ins_2}$\tablefootmark{b}       &1.36$_{-0.46}^{+0.70}$&1.13$_{-0.31}^{+0.48}$&1.14$_{-0.32}^{+0.47}$\\
       $\Gamma$&1.56$_{-*}^{+0.20}$&1.75$_{-0.24}^{+0.17}$&1.95$_{-0.31}^{+0.32}$\\
       $C _{Tor}$&---&---&0.49$_{-0.28}^{+0.47}$\\
       $\theta\rm _{Inc}$ &---&---&63$_{-16}^{+9}$\\
       N$\rm _{H,z}$ &3.95$_{-1.91}^{+*}$&1.95$_{-0.28}^{+0.41}$&2.34$_{-0.58}^{+*}$\\
       N$\rm _{H,S}$ &1.25$_{-0.60}^{+0.24}$&3.74$_{-1.37}^{+3.74}$&4.26$_{-2.54}^{+18.43}$\\
       $f_s$ 10$^{-2}$&0.23$_{-0.13}^{+0.26}$&0.07$_{-*}^{+0.25}$&0.03$_{-*}^{+0.14}$\\
       F$_{2-10{\rm keV}}$ &4.36$_{-2.82}^{+129.64}$&4.56$_{-2.10}^{+1.73}$&4.50$_{-*}^{+7.26}$\\
       F\tablefootmark{c}$_{10-50{\rm keV}}$ &1.15$_{-*}^{+1.65}$&1.12$_{-0.52}^{+0.06}$&1.11$_{-0.64}^{+0.08}$\\
       L$_{2-10{\rm keV}}$ &9.48$_{-5.71}^{+11.06}$&5.33$_{-2.82}^{+4.14}$&11.29$_{-*}^{+27.28}$\\
       L$_{10-50{\rm keV}}$ &19.10$_{-11.50}^{+22.31}$&8.04$_{-4.26}^{+6.23}$&12.33$_{-*}^{+29.80}$\\
       \hline
	\hline
	\vspace{0.02cm}
\end{tabular}

\tablefoot{We summarise here the best-fits of joint \swx--\NuSTAR\ spectra using different torus models at 0.8-50 keV, referred in Section \ref{sec:NGC2788A}. The statistics and degrees of freedom for each fit are also reported. The parameters are reported as in Table \ref{Table:best-fit_MCG-02-12-017} if not mentioned otherwise.}\\
\tablefoottext{a}{$C_{Ins_1}$ = $C_{FPMA/XRT}$ is the cross calibration constant between \NuSTAR\ observation of 2019 and \swx\/.}
\tablefoottext{b}{$C_{Ins_2}$ = $C_{FPMA/XRT}$ is the cross calibration constant between \NuSTAR\ observation of 2020 and \swx\/.}
\tablefoottext{c}{Flux between 10--50\,keV in $10^{-11}$ erg cm$^{-2}$ s$^{-1}$.}

\end{table}

\begin{table}
\renewcommand*{\arraystretch}{1.5}
\centering
\caption{Summary of best-fit solutions of XMM-Newton and NuSTAR data using different models for NGC 1106}
\label{Table:best-fit_NGC1106}
\vspace{.1cm}
   \begin{tabular}{ccccccc}
       \hline
       \hline       
       Model&MyTorus&MyTorus&borus02\\
       &Edge-on&Face-on&&\\
    \hline
       $\chi^2$/dof&357/295&310/295&304/293\\
       $C_{Ins_1}$\tablefootmark{a}       &0.75$_{-0.14}^{+0.15}$&1.05$_{-0.19}^{+0.22}$&1.07$_{-0.19}^{+0.24}$\\
       $C_{Ins_2}$\tablefootmark{b}       &0.92$_{-0.17}^{+0.15}$&1.28$_{-0.22}^{+0.27}$&1.29$_{-0.23}^{+0.28}$\\
       $\Gamma$&1.40$_{-*}^{+0.00}$&1.68$_{-0.22}^{+0.11}$&1.92$_{-0.35}^{+0.44}$\\
       $C _{Tor}$&---&---&0.87$_{-0.24}^{+0.11}$\\
       $\theta\rm _{Inc}$ &---&---&37$_{-9}^{+37}$\\
       N$\rm _{H,z}$ &4.00$_{-0.65}^{+1.73}$&3.43$_{-0.76}^{+*}$&4.79$_{-1.96}^{+*}$\\
       N$\rm _{H,S}$ &1.46$_{-0.12}^{+0.18}$&7.98$_{-3.48}^{+*}$&4.83$_{-1.38}^{+*}$\\
       $f_s$ 10$^{-2}$&0.71$_{-0.19}^{+0.26}$&0.72$_{-0.36}^{+0.71}$&0.46$_{-0.38}^{+1.04}$\\
       kT &0.97$_{-0.11}^{+0.17}$&1.01$_{-0.11}^{+0.47}$&0.99$_{-0.12}^{+0.32}$\\
       kT &0.38$_{-0.31}^{+0.57}$&0.42$_{-0.10}^{+0.21}$&0.42$_{-0.11}^{+0.21}$\\
       F$_{2-10{\rm keV}}$ &2.73$_{-2.70}^{+94.62}$&2.63$_{-2.22}^{+1.87}$&2.60$_{-2.45}^{+11.20}$\\
       F$_{10-50{\rm keV}}$ &7.62$_{-7.61}^{+23.18}$&7.83$_{-7.59}^{+3.02}$&7.98$_{-7.97}^{+11.97}$\\
       L$_{2-10{\rm keV}}$ &6.87$_{-1.66}^{+1.91}$&3.68$_{-0.95}^{+1.58}$&5.02$_{-3.97}^{+5.75}$\\
       L$_{10-50{\rm keV}}$ &18.04$_{-4.37}^{+5.01}$&6.16$_{-1.60}^{+2.65}$&5.50$_{-4.35}^{+6.29}$\\
       \hline
	\hline
	\vspace{0.02cm}
\end{tabular}

\tablefoot{We summarise here the best-fits of joint \XMM--\NuSTAR\ spectra using different torus models at 0.6-50 keV, referred in Section \ref{sec:NGC1106}. The statistics and degrees of freedom for each fit are also reported.The parameters are reported as in Table \ref{Table:best-fit_MCG-02-12-017} if not mentioned otherwise.}
\tablefoottext{a}{$C_{Ins}$ = $C_{FPMA/PN}$ is the cross calibration constant between \NuSTAR\ observation of 2020 and \XMM\ observation of 2019.}\\
\tablefoottext{b}{$C_{Ins}$ = $C_{FPMA/PN}$ is the cross calibration constant between \NuSTAR\ observation of 2020 and \XMM\ observation of 2019.}\\

\end{table}

\begin{table}[hbt!]
\renewcommand*{\arraystretch}{1.5}
\centering
\caption{Summary of best-fit solutions of Chandra and NuSTAR data using different models for ESO406-G004}
\label{Table:best-fit_ESO406-G004}
\vspace{.1cm}
   \begin{tabular}{ccccccc}
       \hline
       \hline       
       Model&MyTorus&MyTorus&borus02\\
       &Edge-on&Face-on&&\\
    \hline
       C-Stat/dof&84/86&84/86&83/84\\
       $C_{Ins_1}$\tablefootmark{a}       &0.73$_{-0.30}^{+0.65}$&0.85$_{-0.36}^{+0.63}$&0.68$_{-0.07}^{+0.60}$\\
       $C_{Ins_2}$\tablefootmark{b}       &0.35$_{-0.15}^{+0.32}$&0.41$_{-0.18}^{+0.31}$&0.34$_{-0.06}^{+0.35}$\\
       $\Gamma$&1.48$_{-*}^{+0.30}$&1.40$_{-*}^{+0.62}$&1.42$_{-*}^{+0.02}$\\
       $C _{Tor}$&---&---&0.10$_{-*}^{+*}$\\
       $\theta\rm _{Inc}$&---&---&18$_{-*}^{+69}$\\
       N$\rm _{H,z}$ &0.85$_{-0.23}^{+0.43}$&0.73$_{-0.14}^{+0.27}$&0.79$_{-0.12}^{+0.04}$\\
       N$\rm _{H,S}$ &0.49$_{-0.36}^{+1.11}$&1.30$_{-1.19}^{+0.73}$&4.97$_{-4.80}^{+*}$\\
       $f_s$ 10$^{-2}$ &0.65$_{-0.62}^{+1.33}$&0.81$_{-*}^{+1.29}$&0.37$_{-*}^{+0.65}$\\
       kT&0.51$_{-*}^{+0.20}$&0.52$_{-*}^{+0.19}$&0.52$_{-0.18}^{+0.18}$\\
       F$_{2-10{\rm keV}}$ &3.41$_{-3.41}^{+1.37}$&3.19$_{-3.19}^{+5.52}$&3.56$_{-3.56}^{+3.47}$\\
       F$_{10-50{\rm keV}}$ &3.50$_{-3.50}^{+0.19}$&3.61$_{-*}^{+10.58}$&3.56$_{-3.56}^{+5.54}$\\
       L$_{2-10{\rm keV}}$ &6.77$_{-3.87}^{+8.68}$&3.97$_{-1.71}^{+26.51}$&6.47$_{-0.37}^{+0.42}$\\
       L$_{10-50{\rm keV}}$ &15.58$_{-8.91}^{+19.96}$&10.27$_{-4.44}^{+68.51}$&16.55$_{-0.93}^{+1.07}$\\
       \hline
	\hline
	\vspace{0.02cm}
\end{tabular}

\tablefoot{We summarise here the best-fits of joint \cha--\NuSTAR\ spectra using different torus models at 0.7-50 keV, referred in section \ref{sec:ESO406-G004}. The statistics and degrees of freedom for each fit are also reported. The parameters are reported as in Table \ref{Table:best-fit_MCG-02-12-017} if not mentioned otherwise.}\\
\tablefoottext{a}\footnote{$C_{Ins_1}$ = $C_{FPMA/ACIS}$ is the cross calibration constant between \NuSTAR\ observation of 2016 and \cha\ observation of 2012.}\\
\tablefoottext{b}{$C_{Ins_2}$ = $C_{FPMA/ACIS}$ is the cross calibration constant between \NuSTAR\ observation of 2020 and \cha\ observation of 2012.}\\

\end{table}


\begin{table}
\renewcommand*{\arraystretch}{1.5}
\centering
\caption{Summary of best-fit solutions of \XMM\/, \cha\ and NuSTAR data using different models for 2MASX J20145928+2523010}
\label{Table:best-fit_2MASXJ20145928+2523010}
\vspace{.1cm}
   \begin{tabular}{ccccccc}
       \hline
       \hline       
       Model&MyTorus&MyTorus&borus02\\
       &Edge-on&Face-on&&\\
    \hline
       $\chi^2$/dof&545/581&545/581&564/580\\
       $C_{Ins_1}$\tablefootmark{a}       &1.98$_{-0.15}^{+0.16}$&1.98$_{-0.15}^{+0.16}$&1.98$_{-0.15}^{+0.15}$\\
       $C_{Ins_2}$\tablefootmark{b}       &1.32$_{-0.08}^{+0.08}$&1.32$_{-0.08}^{+0.08}$&1.31$_{-0.08}^{+0.08}$\\
       $C_{Ins_3}$\tablefootmark{c}       &1.95$_{-0.15}^{+0.16}$&1.95$_{-0.15}^{+0.16}$&1.94$_{-0.15}^{+0.14}$\\
       $\Gamma$&1.79$_{-0.10}^{+0.10}$&1.77$_{-0.08}^{+0.11}$&1.52$_{-0.05}^{+0.04}$\\
       $C _{Tor}$&---&---&1.00$_{-0.23}^{+*}$\\
       $\theta\rm _{Inc}$ &---&---&18$_{-*}^{+*}$\\
       N$\rm _{H,z}$\tablefootmark{d} &2.02$_{-0.16}^{+0.17}$&2.04$_{-0.15}^{+0.19}$&2.18$_{-0.13}^{+0.11}$\\
       N$\rm _{H,S}$\tablefootmark{e} &11.93$_{-2.86}^{+3.85}$&20.15$_{-6.33}^{+8.51}$&17.63$_{-4.75}^{+4.05}$\\
       $f_s$ 10$^{-2}$&0.68$_{-0.39}^{+0.40}$&0.42$_{-*}^{+0.44}$&1.06$_{-0.43}^{+0.40}$\\
       F\tablefootmark{f}$_{2-10{\rm keV}}$ &1.83$_{-0.06}^{+0.04}$&1.83$_{-0.08}^{+0.05}$&1.81$_{-0.09}^{+0.04}$\\
       F$_{10-50{\rm keV}}$ &5.04$_{-0.45}^{+0.21}$&5.07$_{-0.52}^{+0.46}$&6.81$_{-0.53}^{+0.41}$\\
       L$_{2-10{\rm keV}}$ &6.27$_{-0.93}^{+1.09}$&5.95$_{-0.89}^{+1.18}$&8.20$_{-0.64}^{+0.70}$\\
       L$_{10-50{\rm keV}}$ &8.76$_{-1.31}^{+1.53}$&8.52$_{-1.27}^{+1.69}$&17.57$_{-1.37}^{+1.50}$\\
       \hline
	\hline
	\vspace{0.02cm}
\end{tabular}

\tablefoot{We summarise here the best-fits of joint \XMM\ , \cha\ and \NuSTAR\ spectra using different torus models at 0.6-50 keV, referred in Section \ref{sec:2MASXJ20145928+2523010}. The statistics and degrees of freedom for each fit are also reported. The parameters are reported as in Table \ref{Table:best-fit_MCG-02-12-017} if not mentioned otherwise.}\\
\tablefoottext{a}{$C_{Ins_1}$ = $C_{ACIS/PN}$ is the cross calibration constant between \cha\ observation of 2018 and \XMM\ observation of 2017.}\\
\tablefoottext{b}{$C_{Ins_2}$ = $C_{FPMA/PN}$ is the cross calibration constant between \NuSTAR\ observation of 2017 and \XMM\ observation of 2017.}\\
\tablefoottext{c}{$C_{Ins_3}$ = $C_{FPMA/PN}$ is the cross calibration constant between \NuSTAR\ observation of 2020 and \XMM\ observation of 2017.}\\
\tablefoottext{d}{``Line of sight'' column density in $10^{22}$\,cm$^{-2}$.}\\
\tablefoottext{e}{Average column density from scattering in $10^{22}$\,cm$^{-2}$.}\\
\tablefoottext{f}{Flux between 2--10\,keV in $10^{-12}$ erg cm$^{-2}$ s$^{-1}$.}\\

\end{table}

\begin{table}
\renewcommand*{\arraystretch}{1.5}
\centering
\caption{Summary of best-fit solutions of XMM-Newton and NuSTAR data using different models for ESO138-G001}
\label{Table:best-fit_ESO138-G001}
\vspace{.1cm}
   \begin{tabular}{ccccccc}
       \hline
       \hline       
       Model&MyTorus&MyTorus&borus02\\
       &Edge-on&Face-on&&\\
    \hline
       $\chi^2$/dof&2636/1818&2566/1818&2419/1816\\
       $C_{Ins_1}$\tablefootmark{a}       &1.15$_{-0.04}^{+0.04}$&1.15$_{-0.05}^{+0.05}$&1.16$_{-0.05}^{+0.03}$\\
       $C_{Ins_2}$\tablefootmark{b}       &1.07$_{-0.04}^{+0.04}$&1.08$_{-0.04}^{+0.04}$&1.09$_{-0.05}^{+0.03}$\\
       $\Gamma$&1.53$_{-0.05}^{+0.05}$&1.66$_{-0.05}^{+0.03}$&1.98$_{-0.03}^{+0.01}$\\
       $C _{Tor}$&---&---&0.80$_{-0.12}^{+0.03}$\\
       $\theta\rm _{Inc}$ &---&---&18$_{-*}^{+*}$\\
       N$\rm _{H,z}$ &0.33$_{-0.02}^{+0.02}$&0.34$_{-0.02}^{+0.01}$&0.47$_{-0.06}^{+0.07}$\\
       N$\rm _{H,S}$ &1.47$_{-0.49}^{+0.60}$&3.00$_{-0.55}^{+0.37}$&10.35$_{-0.08}^{+0.08}$\\
       Fe $K_{\alpha}$\tablefootmark{c}&6.42$_{-0.01}^{+0.00}$&6.42$_{-0.01}^{+0.00}$&6.44$_{-0.01}^{+0.01}$\\
       Fe $K_{\alpha}$ norm\tablefootmark{d}10$^{-5}$ &2.28$_{-0.11}^{+0.10}$&1.91$_{-0.10}^{+0.11}$&1.04$_{-0.13}^{+0.10}$\\
       $f_s$ 10$^{-2}$&7.46$_{-0.94}^{+1.07}$&8.14$_{-0.87}^{+0.67}$&3.22$_{-0.24}^{+0.10}$\\
       kT &0.68$_{-0.01}^{+0.01}$&0.68$_{-0.01}^{+0.01}$&0.74$_{-0.01}^{+0.01}$\\
       F\tablefootmark{e}$_{2-10{\rm keV}}$ &2.25$_{-0.07}^{+0.04}$&2.24$_{-0.05}^{+0.04}$&2.24$_{-0.27}^{+0.11}$\\
       F\tablefootmark{f}$_{10-50{\rm keV}}$ &1.42$_{-0.11}^{+0.03}$&1.44$_{-0.09}^{+0.03}$&1.43$_{-0.09}^{+0.06}$\\
       L$_{2-10{\rm keV}}$\tablefootmark{g} &11.59$_{-1.88}^{+2.25}$&12.11$_{-2.04}^{+0.95}$&3.61$_{-0.94}^{+0.30}$\\
       L$_{10-50{\rm keV}}$\tablefootmark{h} &24.51$_{-3.97}^{+4.75}$&20.90$_{-3.52}^{+1.65}$&3.34$_{-0.87}^{+0.28}$\\
       \hline
	\hline
	\vspace{0.02cm}
\end{tabular}

\tablefoot{We summarise here the best-fits of joint \XMM--\NuSTAR\ spectra using different torus models at 0.6-50 keV, referred in Section \ref{sec:ESO138-G001}. The statistics and degrees of freedom for each fit are also reported. The parameters are reported as in Table \ref{Table:best-fit_MCG-02-12-017} if not mentioned otherwise.}\\
\tablefoottext{a}{$C_{Ins_1}$ = $C_{FPMA/PN}$ is the cross calibration constant between \NuSTAR\ observation of 2016 and \XMM\ observation of 2013.}\\
\tablefoottext{b}{$C_{Ins_2}$ = $C_{FPMA/PN}$ is the cross calibration constant between \NuSTAR\ observation of 2020 and \XMM\ observation of 2013.}\\
\tablefoottext{c}{Energy of the Iron $K_{\alpha}$ line in keV.}\\
\tablefoottext{d}{Normalization of line component depicting total photons in cm$^{-2}$\,s$^{-1}$.}\\
\tablefoottext{e}{Flux between 2--10\,keV in $10^{-12}$ erg cm$^{-2}$ s$^{-1}$.}\\
\tablefoottext{f}{Flux between 10--50\,keV in $10^{-11}$ erg cm$^{-2}$ s$^{-1}$.}\\
\tablefoottext{g}{Intrinsic luminosity between 2--10\,keV in $10^{41}$ erg s$^{-1}$.}\\
\tablefoottext{h}{Intrinsic luminosity between 2--10\,keV in $10^{41}$ erg s$^{-1}$.}\\

\end{table}

\clearpage

\section{Figures of X-Ray Spectra}
\label{appendix2:figures}


\begin{figure}[hbt!]
\begin{center}
\includegraphics[scale = 1.0, width = 18 cm, trim = 10 520 10 0, clip]{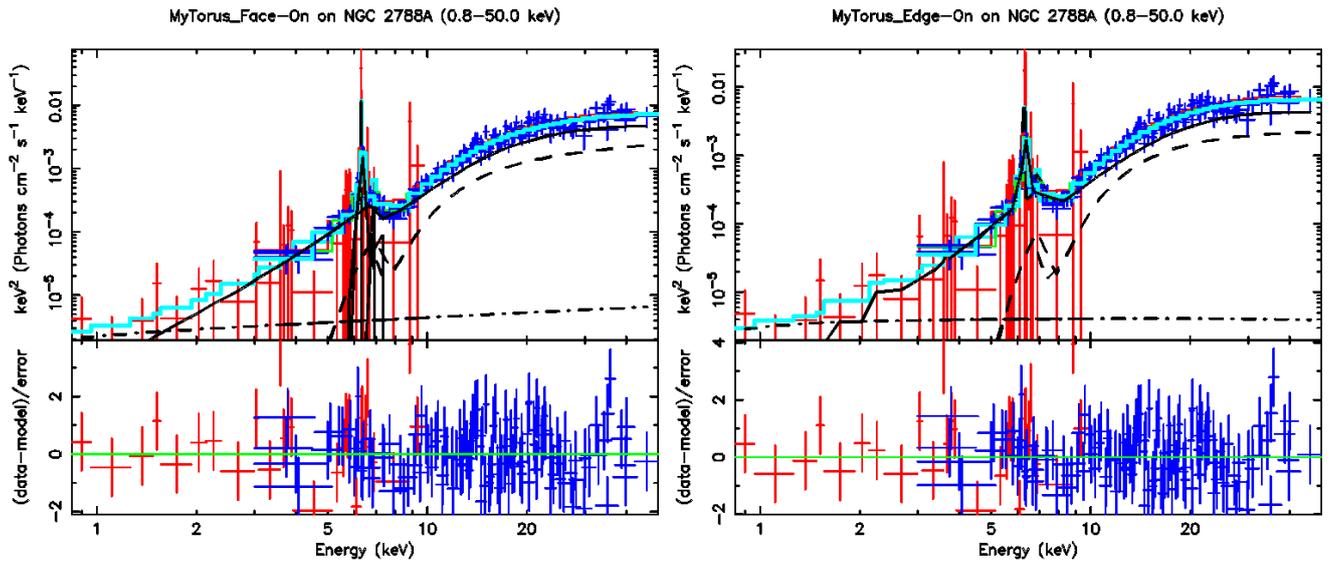} 
\caption{Same as Figure \ref{fig:MCG-02-12-017}, for NGC 2788A, without mekal.}
\label{fig:NGC2788A}
\end{center}
\end{figure}


\begin{figure}[hbt!]
\begin{center}
\includegraphics[scale = 1.0, width = 18 cm, trim = 10 520 10 0, clip]{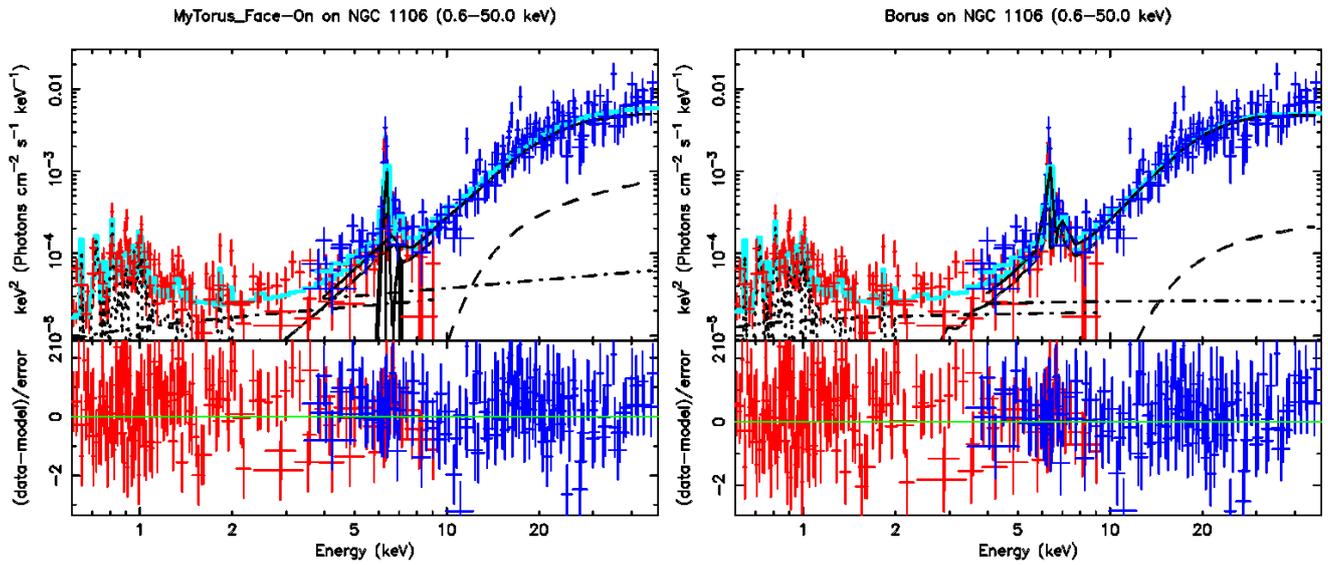} 
\caption{Same as Figure \ref{fig:MCG-02-12-017}, for NGC 1106.}
\label{fig:NGC1106}
\end{center}
\end{figure}


\begin{figure*}[hbt!]
\begin{center}
\includegraphics[scale = 1.0, width = 18 cm, trim = 10 520 10 0, clip]{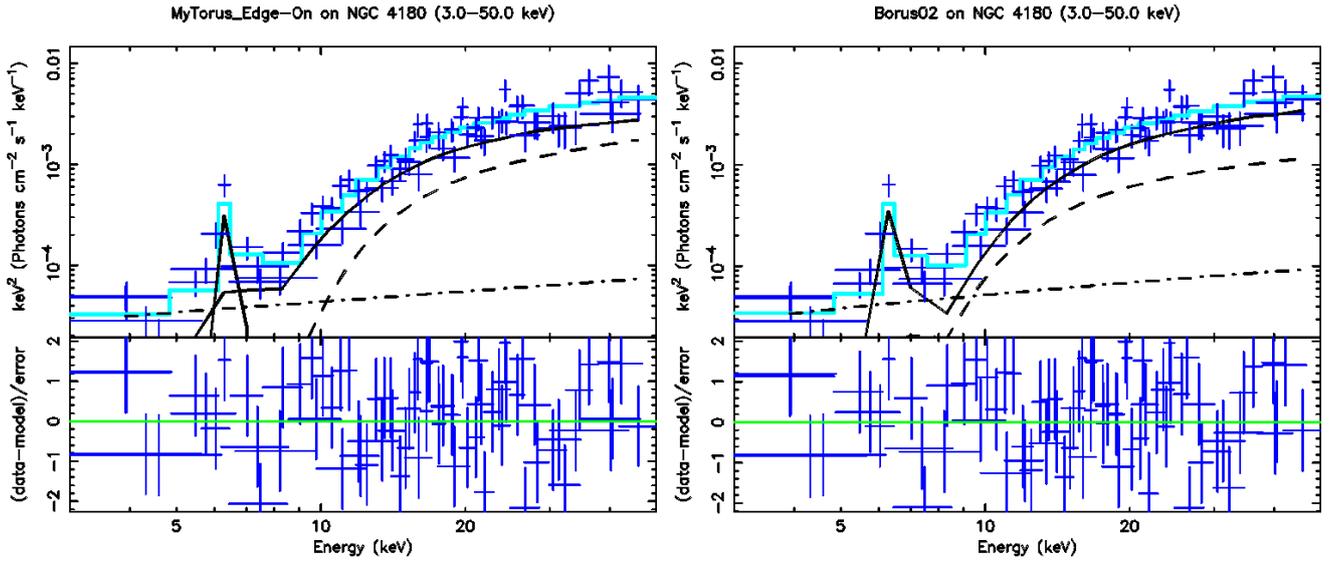} 
\caption{Same as Figure \ref{fig:MCG-02-12-017}, for NGC 4180, without any soft X-ray points and mekal.}
\label{fig:NGC4180}
\end{center}
\end{figure*}


\begin{figure*}[hbt!]
\begin{center}
\includegraphics[scale = 1.0, width = 18 cm, trim = 10 520 10 0, clip]{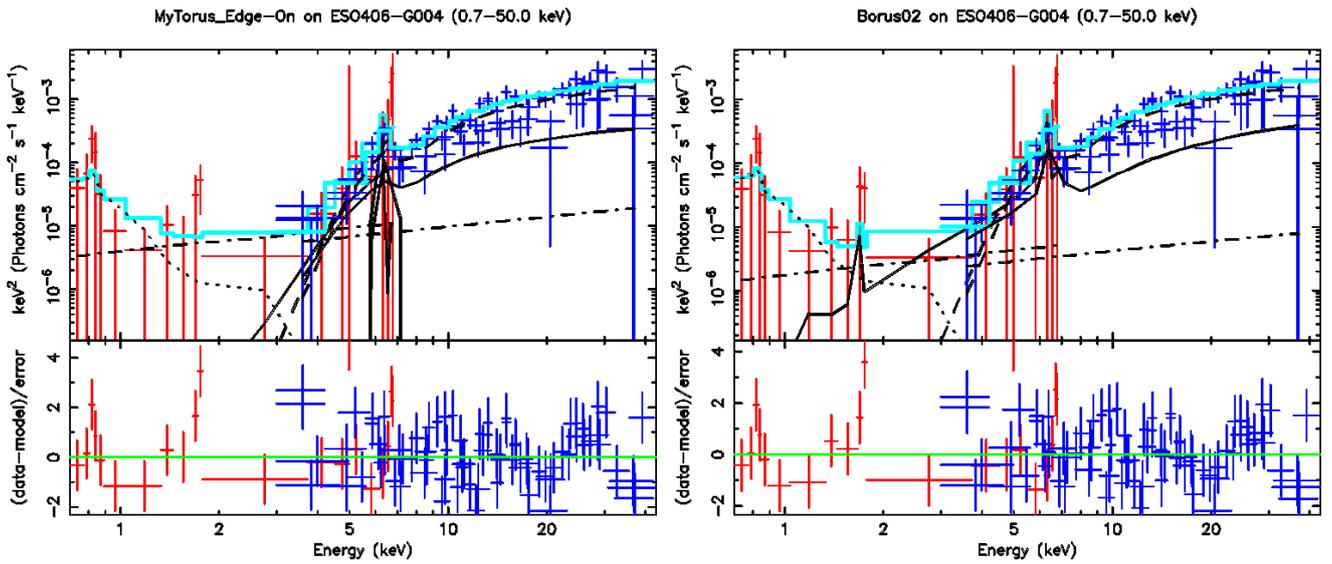} 
\caption{Same as Figure \ref{fig:MCG-02-12-017}, for ESO406-G004.}
\label{fig:ESO406-G004}
\end{center}
\end{figure*}


\begin{figure*}[hbt!]
\begin{center}
\includegraphics[scale = 1.0, width = 18 cm, trim = 10 520 10 0, clip]{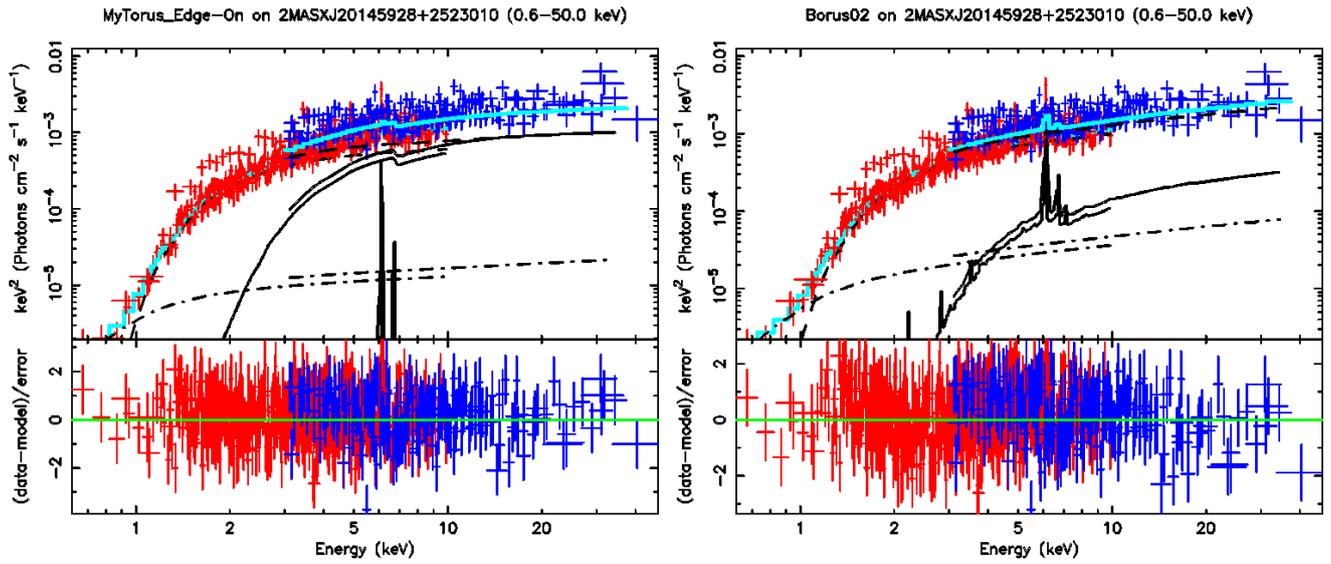} 
\caption{Same as Figure \ref{fig:MCG-02-12-017}, for 2MASXJ20145928+2523010, without mekal.}
\label{fig:2MASXJ20145928+2523010}
\end{center}
\end{figure*}


\begin{figure*}[hbt!]
\begin{center}
\includegraphics[scale = 1.0, width = 18 cm, trim = 10 520 10 0, clip]{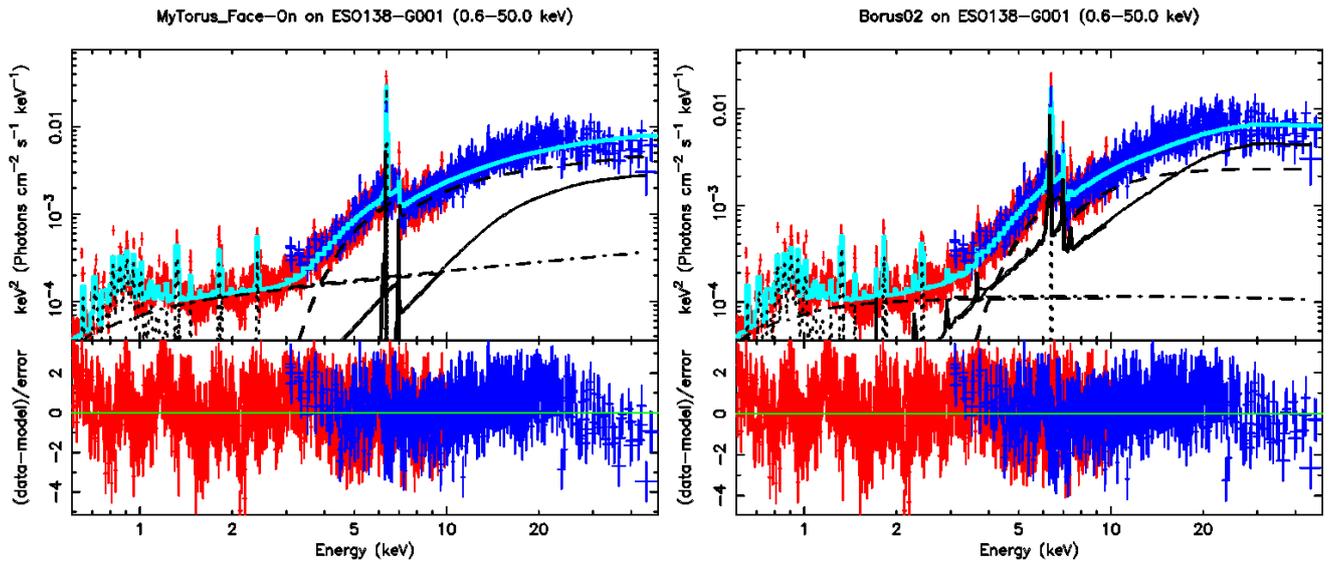} 
\caption{Same as Figure \ref{fig:MCG-02-12-017}, for ESO138-G001, with extra four Gaussian line profiles (dot).}
\label{fig:ESO138-G001}
\end{center}
\end{figure*}

\end{appendix}
\end{document}